\documentclass[aps,prl,reprint,groupedaddress,amsmath, amssymb, showpacs]{revtex4-1}

\usepackage{graphicx}
\usepackage{dcolumn}
\usepackage{bm}
\usepackage{amsmath,amssymb}
\usepackage{pstool}

\usepackage[babel=true]{csquotes}

\usepackage{newfloat,algcompatible}
\usepackage[size=small]{caption}
\usepackage{etoolbox}

\AtEndEnvironment{algorithm}{\noindent\hrulefill\par\nobreak\vskip-8pt}

\DeclareFloatingEnvironment[
    fileext=loa,
    listname=List of Algorithms,
    name=ALGORITHM,
    placement=tbhp,
]{algorithm}
\DeclareCaptionFormat{algorithms}{\vskip-15pt\hrulefill\par#1#2#3\vskip-6pt\hrulefill}
\algblock[MyInput]{MyInput}{MyEndInput}
\algblockdefx[MyInput]{MyInput}{MyEndInput}%
    [1]{\textbf{Input} #1}%

\RequirePackage[normalem]{ulem}
\RequirePackage{color}\definecolor{RED}{rgb}{1,0,0}\definecolor{BLUE}{rgb}{0,0,1}

\usepackage{subfig}
\newcommand{\ie}{i.e.\ }
\newcommand{\etal}{et al.\ }
\newcommand{\eg}{e.g.\ }
\newcommand{\etc}{etc.\ }
\usepackage{color}
\begin{document}


\title{Fast community detection using local neighbourhood search}


\author{Arnaud Browet}
\email[Corresponding author:]{arnaud.browet@uclouvain.be}

\author{P.-A. Absil}

\author{Paul Van Dooren}
\affiliation{
	Universit\'{e} catholique de Louvain\\
	Department of Mathematical Engineering, Institute of Information and Communication Technologies, Electronics and Applied Mathematics (ICTEAM)\\
	Av. Georges Lema\^{i}tre 4-6, B-1348 Louvain-la-Neuve, Belgium.
}


\date{\today}

\begin{abstract}
Communities play a crucial role to describe and analyse modern networks. However, the size of those networks has grown tremendously with the increase of computational power and data storage. While various methods have been developed to extract community structures, their computational cost or the difficulty to parallelize existing algorithms make partitioning real networks into communities a challenging problem. In this paper, we propose to alter an efficient algorithm, the Louvain method, such that communities are defined as the connected components of a tree-like assignment graph. Within this framework, we precisely describe the different steps of our algorithm and demonstrate its highly parallelizable nature. We then show that despite its simplicity, our algorithm has a partitioning quality similar to the original method on benchmark graphs and even outperforms other algorithms. We also show that, even on a single processor, our method is much faster and allows the analysis of very large networks.
\end{abstract}


\maketitle

\section{Introduction}
Over the years, the evolution of information and communication technology in science and industry has had a significant impact on how collected data are being used to understand and analyse complex phenomena \cite{halevy2009}. With the increase in storage capacity and computational power, the amount of collected data has grown tremendously. Many researchers believe nowadays that clever analysis of those so-called Big Data are at the core of many successful projects and that suitable algorithms must be developed to deal with such data \cite{torralba2008, skillicorn2012}.
In this context, graph theory has provided powerful tools to identify relevant patterns in the data. Indeed, the content of such large databases can often be expressed as a set of agents, represented as nodes, associated with their mutual interactions, represented as edges, revealing a network structure \cite{Albert2007, lazer2009}. Examples are telecommunication networks, with calling and texting habits revealing interactions between mobile phone users \cite{Onnela2007}, online social networks of friendships \cite{Kumpula2009}, or the World Wide Web \cite{Albert1999}.
As mentioned already, a common attribute of modern networks is their fast growing size. The mobile phone penetration in the world is close to $85\%$ with an estimate of $6$ billions devices around the globe, the largest online social network now records more than $1$ billion users and the size of the web was recently estimated at about $40$ billions pages.
These networks are too large to comprehend and even a simple visualization of the network is often impossible. 

To establish some behavioural properties about the objects of interest, a popular technique is to cluster together highly similar nodes. When the pairwise node similarity is encoded in the edge weight, this task is known in graph theory as community detection and has been introduced by Newman \cite{Newman2000}. Community detection has already proven to be of great interest in many different research areas such as epidemiology \cite{Tunk2012}, influence and spread of information over social networks \cite{Wu2008, Liu2005}, analysis of the air transportation network \cite{Guimera2005a}, detection of roles in metabolic networks \cite{Guimera2005}, image processing \cite{Browet2011}, \etc There is no universally accepted definition of communities. However, they are informally defined as sets of nodes with high internal density, either in the number of internal edges or their weight, and low external density with the rest of the network. Hence, to extract community structures, various cost functions have been proposed based on suitable rewards for edges that are present or penalties for missing edges within each community. These cost functions are in general optimized over the community assignment of each node. Under some cost symmetry assumptions, a general class of cost functions has been proposed by Reichardt and Bornholdt \cite{Reichardt2006} as 
\begin{equation}
Q_{RB} = \sum\limits_{i,j=1}^{N} \left( W_{ij} - \gamma_{RB}p_{ij}\right)\delta\left(\sigma_{i},\sigma_{j}\right).
\label{eq:QRB}
\end{equation}
Here, $W$ is the weighted adjacency matrix of the graph of $N$ nodes, $p_{ij}$ is the expected weight of an edge between nodes $i$ and $j$ known as the random null model and $\gamma_{RB}$ is a resolution parameter. In this model, $\sigma_i$ denotes the community index of node $i$ and $\delta\left(\sigma_i,\sigma_j\right) = 1$ if $\sigma_i = \sigma_j$ and 0 otherwise.
One of the most popular cost functions for community detection is the modularity introduced by Newman \& Girvan \cite{Newman2004, Newman2004a, Leicht2007} which can be obtained by choosing $\gamma_{RB}=1$ and $p_{ij} = \frac{k_i^{out}k_j^{in}}{m}$, where $m$ is the sum of the weight of all the edges in the graph and $k_i^{in/out}$ is the incoming (respectively outgoing) strength of node $i$. Modularity compares the actual edge density within each community to the expected edge density in a random network with the same weighted degree distribution. This cost function has been widely used in various fields \cite{Guimera2005a,Kashtan2005,Mucha2010,Conover2013} but recent studies have shown that it suffers from a resolution limit \cite{Fortunato2007,Schaub2012}, \ie the size of the communities extracted based on the optimization of modularity depends on the size of the network. To avoid this undesirable property, other cost functions have been defined such as the constant Potts model \cite{Reichardt2006,Traag2011} where the expected weight $p_{ij} = \gamma$ produces communities whose sizes are independent of the scale of the network and for which the optimal partition of the entire graph is also optimal for any sub-graph induced by a set of communities. While this cost function provides in general a better partition of the network, it might require to also optimize the parameter $\gamma$ to extract the most significant partition \cite{Traag2011,Delvenne2010,Lambiotte2008, Traag2013}.
Another efficient cost function, inspired by information theory, has been defined by Rosvall \& Bergstrom \cite{Rosvall2008} and is based on the compression of the description length of the average path of a random walker over the network. 
An exhaustive analysis of the existing cost functions to detect communities in networks is out of the scope of this paper but detailed reviews have been provided by Fortunato \cite{Fortunato2010} and Porter \etal \cite{Porter2009}.\\

Finding the optimal partition for a given cost function is in general a difficult problem. For example, maximizing the modularity has been proven to be NP-hard \cite{Brandes2008}. Hence, different algorithms have been developed to approximate the optimal partition of a network. All the existing heuristics designed to extract community structures have to balance the quality of the partition with respect to the time complexity of the algorithm. In this paper, we propose a synchronous version of the so-called Louvain method developed by Blondel \etal \cite{Blondel2008}. We show that the communities extracted by our algorithm are of similar quality but with a much smaller time complexity as well as being highly synchronizable on a parallel architecture. In the next sections, we first briefly review the original algorithm of Blondel \etal and present each step of our hierarchical procedure. We then assess the quality of the extracted clusters and the average computational time required by our algorithm and some other popular methods on benchmark graphs. We show that our algorithm outperforms the 
other methods in terms of computational complexity while producing communities of similar quality.

\section{Review of the Louvain method}
\label{sec:algorithm}
Our algorithm is designed to iteratively create community structures by using simple rules, largely inspired by the so-called Louvain method \cite{Blondel2008}. The Louvain method is a hierarchical clustering algorithm divided in two phases. In the first phase, communities are initialized such that each node defines its own community. Then, based on a chosen cost function, individual nodes are sequentially moved to one of their neighbouring communities if that produces a positive gain of the cost function. We call this a \emph{correction step}. For example, using the modularity cost function, one can compute the gain of moving a node $i$ to a community $c$ as 
\begin{equation}
Q_M({i\rightarrow c})= \sum\limits_{j\in c} \left(W_{ij}+W_{ji}\right) - \frac{k_i^{out}s_c^{in}}{m} - \frac{k_i^{in}s_c^{out}}{m},
\label{eq:modularity_gain}
\end{equation}
where $s_c^{out/in}$ is the sum of the outgoing/incoming strength of nodes in community $c$, $s_c^{out/in} = \sum\limits_{j\in c}k_j^{out/in}$.

As a result, computing the gain to assign a node to another community is an inexpensive operation if the total strength of each community is stored and updated after each correction step.
It is worth mentioning that node assignments are often reconsidered so that it is possible to remove a node from its community and reassign it to another community if that produces a positive gain of the cost function according to the current community distribution.

In a second phase, when no correction with a strictly positive gain can be found for any node in the network, the graph is collapsed to create a new network based on the communities built in the first phase. Each community is aggregated into a single node and the edges are aggregated such that nodes in the collapsed graph are linked by weighted edges (including self loops) representing the sum of the edges between the associated communities in the clustered graph. Formally, a graph represented by its weighted adjacency matrix $W\in \mathbb{R}^{n\times n}$ and a community matrix $C \in \left\{0,1\right\}^{k\times n}$, such that $C_{ij} = 1$ if community $i$ contains node $j$, is collapsed to produce a new network with $k$ nodes represented by a weighted adjacency matrix 
\begin{equation}
	\overline{W} = C\,WC^T
\label{eq:collapsed_adjacency}
\end{equation}
We call this an \emph{aggregation step}.\\

These two phases are then recursively applied to the collapsed graphs until no community can be found, \ie $C$ is the identity matrix in the last step. Hence, the algorithm defines potentially multiple hierarchical levels of clustering. The finest level corresponds to a state where each node defines its own community and each level of the partition is composed of an aggregation of the communities at the previous level.

While this algorithm has proven to produce good community structures \cite{Blondel2008}, it might require much time to create the output partition of large networks because of the sequential correction steps. Indeed, each node might be considered many times before no more improvements can be found for the partition. Those sequential correction steps also make the algorithm hard to parallelize on a multiple core architecture, making the analysis of very large networks computationally challenging.

\begin{algorithm}[t]
\caption{Synchronized Louvain}
\label{algo_sync_louvain}
\begin{algorithmic}[1]
\REQUIRE{$G_0(V_0,E_0)$ a graph}\vskip5pt
	\State{$k=0,\, G = G_0$}\vskip3pt
	
	\REPEAT
		\State $\forall\, i\in V_k,\, a_i=Assign(i)$\label{algo_assign}\COMMENT{assignment graph}
		\State $\forall\, i,\, c_i = CC_i(a)$ \Comment{connected components}\vskip5pt
		
		\State $s=\left|V_k\right|$\COMMENT{\# of switched nodes}
		\WHILE{$s>0$}\vskip5pt
		\Statex \hskip10pt \underline{\emph{Positive Correction:}}\vskip3pt
			\WHILE{$\exists i,\,Q(i\rightarrow c_i)<0$}\label{algo_positive_corr_crit}\Comment{negative local gain}\vskip2pt
				
				\State{$a_{c_i} = Split(c_i$)}\label{algo_positive_corr}\vskip2pt
				\State $\forall\,j \in c_i,\, c_j = CC_j(a)$\vskip5pt
			\ENDWHILE\vskip5pt
			
			\Statex \hskip10pt \underline{\emph{Maximal Correction:}}\vskip3pt
			
			\State $s=0$\vskip1pt
			
			\FORALL{$i$}\vskip2pt
				\IF{$Q(i\rightarrow c_i) < Q(i\rightarrow c_j)}$\vskip1pt
						\Comment{non maximal local gain}\vskip2pt
					\State $a_i = Switch(i\rightarrow c_j)$\label{algo_maximal_corr}
					\State $s = s+1$\vskip3pt
				\ENDIF
			\ENDFOR\vskip5pt
			\IF{$s > 0$}
				\State $\forall\,i,\, c_i = CC_i(a)$
			\ENDIF\vskip2pt
		\ENDWHILE\vskip3pt
		 \State $C_k = \left\{c_i\right\}$
		\State $G_{k+1} = Collapse(G_k,C_k)$\label{algo_maximal_collapse}\Comment{community aggregation}
		\State{k=k+1}\vskip1pt
	\UNTIL{$\left|V_k\right| = \left|V_{k-1}\right|$}
	\vskip5pt
\ENSURE{$\left\{C_k\right\}$}
\end{algorithmic}
\end{algorithm}


\section{Simultaneous Local Neighbourhood Search Algorithm}

We propose to alter the correction steps applied in the Louvain algorithm to produce synchronized corrections of the nodes, allowing a high degree of parallelization on multiple processors. The aggregation step has been kept to provide different hierarchical levels of communities. The pseudo-code of our algorithm\footnote{code available at http://sites.uclouvain.be/absil/browet} is presented in Algorithm \ref{algo_sync_louvain} and the next sections highlight the main differences with the original Louvain method.

\subsection{Assignment graph}
 \begin{figure}
	\subfloat[Input graph]{ \includegraphics[width=0.7\columnwidth]{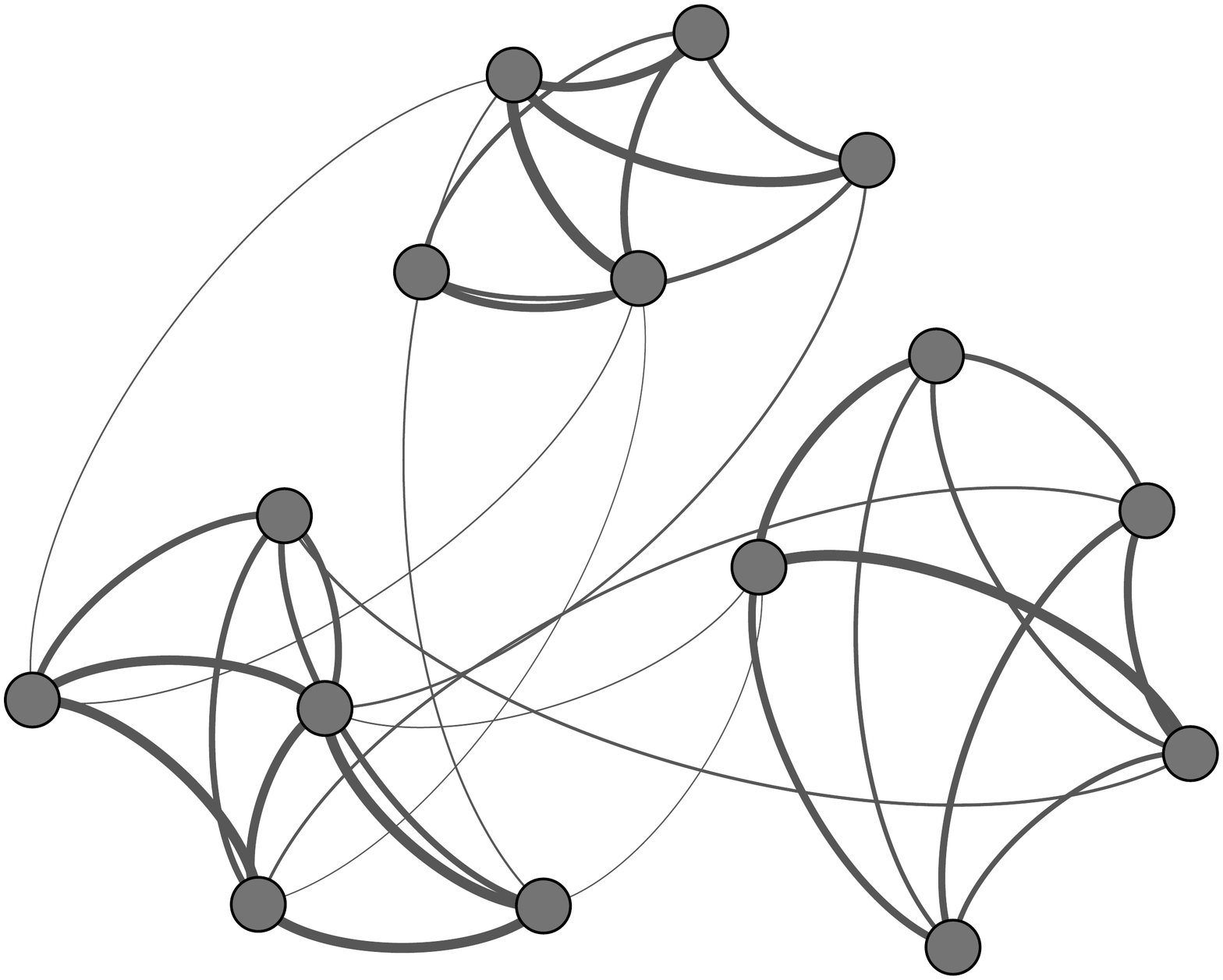}\label{fig:assign_graph_input}}\\
	\subfloat [Assignment graph]{ \includegraphics[width=0.7\columnwidth]{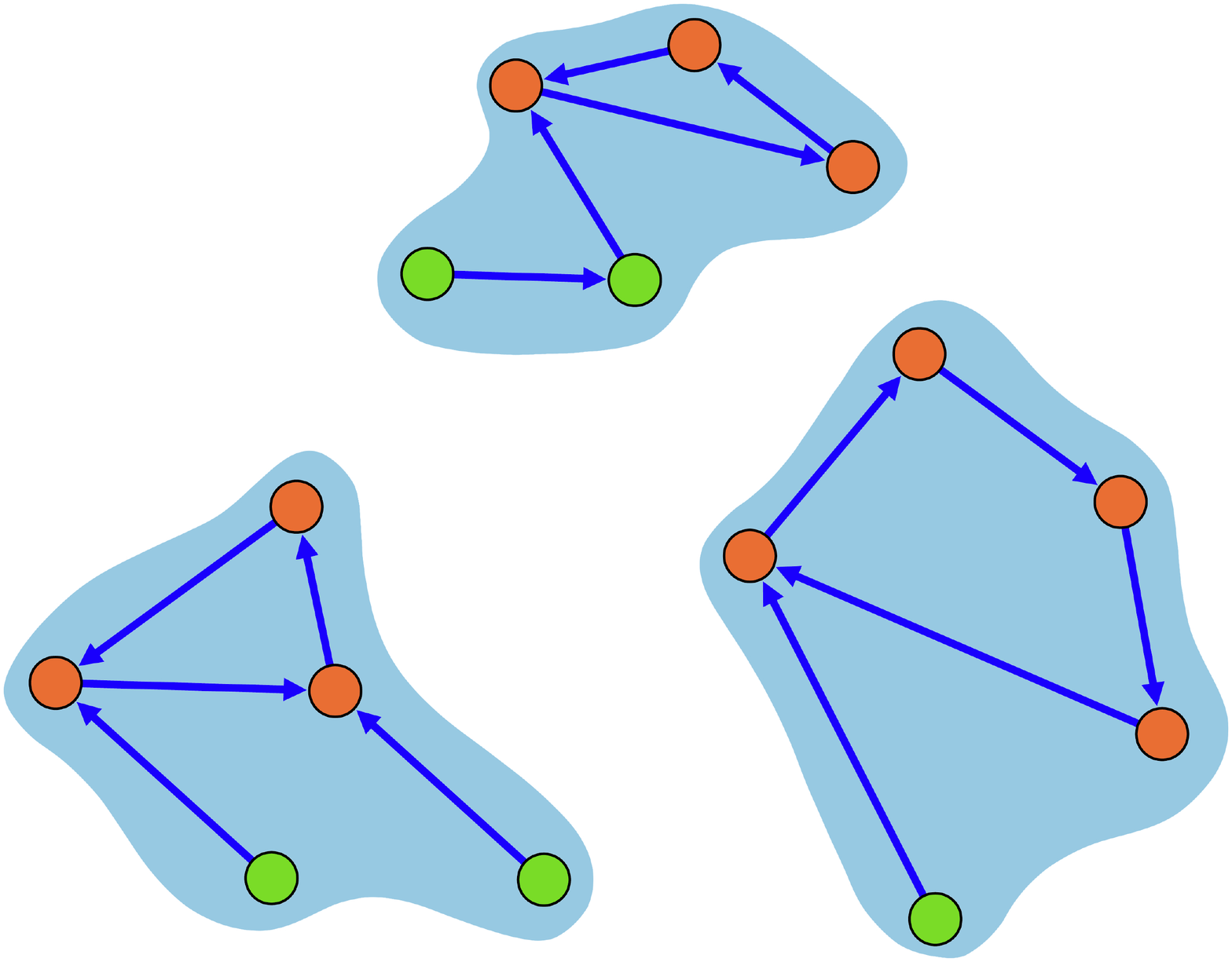}\label{fig:assign_graph_assign}}
 \caption{Based on an input graph (a), the assignment graph, defining the communities, is computed by choosing the best neighbour for each individual node (b) (Color online: orange nodes form the strongly connected component (SCC) and green nodes constitute multiple directed branches leading to the SCC)}
 \label{fig:assign_graph}
 \end{figure}

 \begin{figure*}[t]
 \subfloat[Initial community]{\includegraphics[width=0.4\columnwidth]{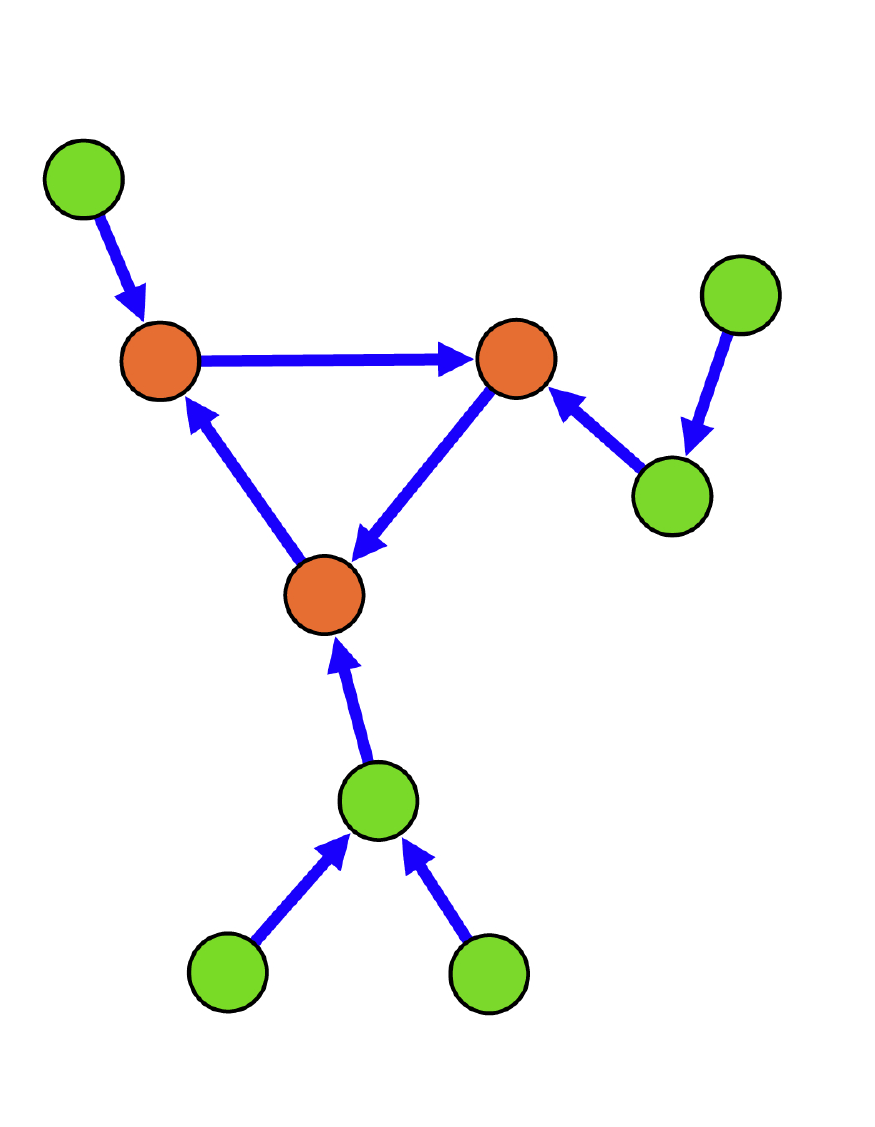}\label{fig:positive_correction_input}}\hspace{1cm}
 \subfloat[Branch split]{\includegraphics[width=0.4\columnwidth]{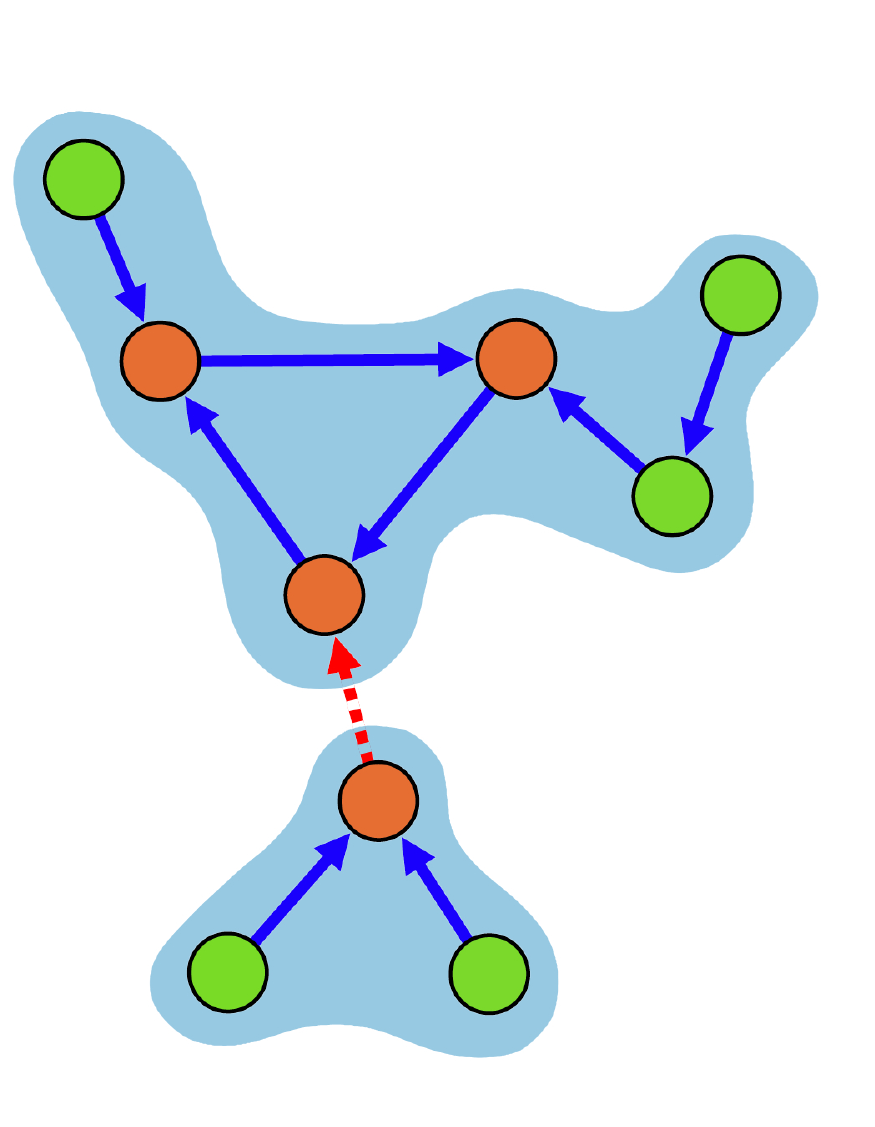}\label{fig:positive_correction_branch}}\hspace{1cm}
 \subfloat[SCC split]{\includegraphics[width=0.4\columnwidth]{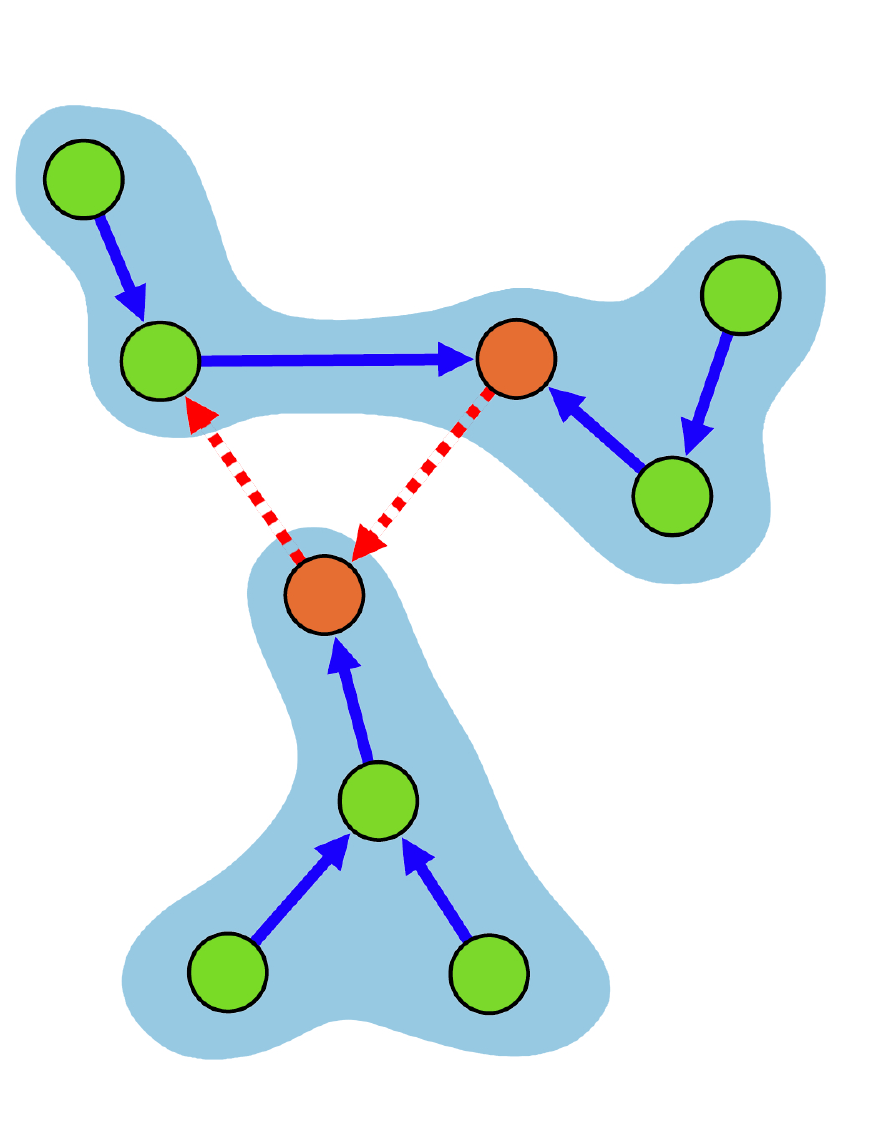}\label{fig:positive_correction_scc}}
   \caption{Positive correction step applied to an input community (a) containing a node with negative local gain. The possible splits are either to remove one assignment within any of the branches (b) or any pair of assignments within the SCC (c). Positive corrections always lead to a bisection of the input community.}
   \label{fig:positive_correction}
 \end{figure*}
 
Our algorithm is designed to build communities spanned by directed tree-like structures covering the entire network. To initialize the community structure, each node is assigned to its best neighbour based on the chosen cost function independently of the choice of the other nodes (function \emph{Assign}, Algorithm \ref{algo_sync_louvain}, step \ref{algo_assign}). For nodes having multiple neighbours that provide the best positive gain, one of them is chosen at random. Using the modularity cost function, the best neighbour of each node can be computed as 
\begin{eqnarray}
	\forall i, a_i = \arg\max_j Q_M\left(i\rightarrow \{j\}\right).
\label{eq:best_gain}
\end{eqnarray}
One can see that each node has a null gain to be assigned to itself, $Q_M\left(i\rightarrow \{i\}\right) = 0$. This implies that the gain to assign each node to its best neighbour must be non negative and all nodes without any neighbour that provides a positive gain will always be self-assigned.
As depicted in Figure \ref{fig:assign_graph_assign}, this assignment step produces the initial structure of our communities defined as the weakly connected components of the assignment graph. Each community contains exactly one directed cycle, the strongly connected component (SCC), defining the core of the community, and a set of directed branches leading to the SCC. The SCC might be of any size, but there is always one and only one SCC within each community because they consist of a directed tree with exactly one additional edge \cite{West2001}. This tree-like structure spanning each community will be maintained through all the phases of our algorithm.
 
While each node had previously chosen the neighbour providing the largest positive gain of the cost function, the synchronous assignments might produce communities in which the presence of some nodes decreases the value of the cost function. As in the Louvain method, this implies that some correction steps might be indicated to increase the overall quality of the communities. Our algorithm considers two types of corrections applied recursively until a local optimum has been reached. We call them the positive and the maximal corrections and we will discuss them in the next sections.

\subsection{Positive correction}
The first type of correction is designed to ensure that every node has at least a positive contribution to the cost function. Indeed, when communities are built synchronously, some nodes might have a negative local gain in the constructed communities. We define the local gain of a node $i$ in its community $c_i$ as the sum of the assignment gains with each of the vertices in this community, \eg using equation (\ref{eq:modularity_gain}), $$Q_M\left(i\rightarrow c_i\right) = \sum\limits_{j\in c_i} Q_M\left(i\rightarrow \{j\}\right).$$
Therefore, for each community containing a node with a negative local gain, $Q_M\left(i\rightarrow c_i\right)<0$, we recursively look for the optimal bisection that maintains the rest of the assignment graph. We call this procedure a positive correction (function \emph{Split} - Algorithm \ref{algo_sync_louvain}, step \ref{algo_positive_corr}).

\noindent Within a community, two different kinds of bisection preserving the rest of the assignment graph can be considered to increase the global value of the cost function, as represented in Figure \ref{fig:positive_correction}. First, the algorithm checks the result of removing any single assignment within one of the directed branches. If there exist multiple bisections that produce a strictly positive gain of the cost function, our algorithm selects the best amongst all the possible removals, effectively splitting the community in two, see Figure \ref{fig:positive_correction_branch}. The node whose assignment is removed is then considered self assigned, which implies that one of the separated community will have a SCC of size $1$.
If no positive correction can be found within any of the branches, our algorithm computes the effect of removing any pair of assignments within the SCC. This does not increase dramatically the computational cost because the SCC's sizes are generally much smaller than the size of the communities. Again, the best pair of assignments to remove will be selected amongst all the strictly positive splits and when such a positive correction is applied, the two newly formed communities contain SCCs of size $1$.

\noindent The positive correction of a community is dominated by a linear complexity in the number of nodes within the branches of a community and is independent of the state of all the other communities, so each split can be assigned to a different processor.

\subsection{Maximal correction}
 \begin{figure*}
 \subfloat[Initial communities]{\includegraphics[width=0.5\columnwidth]{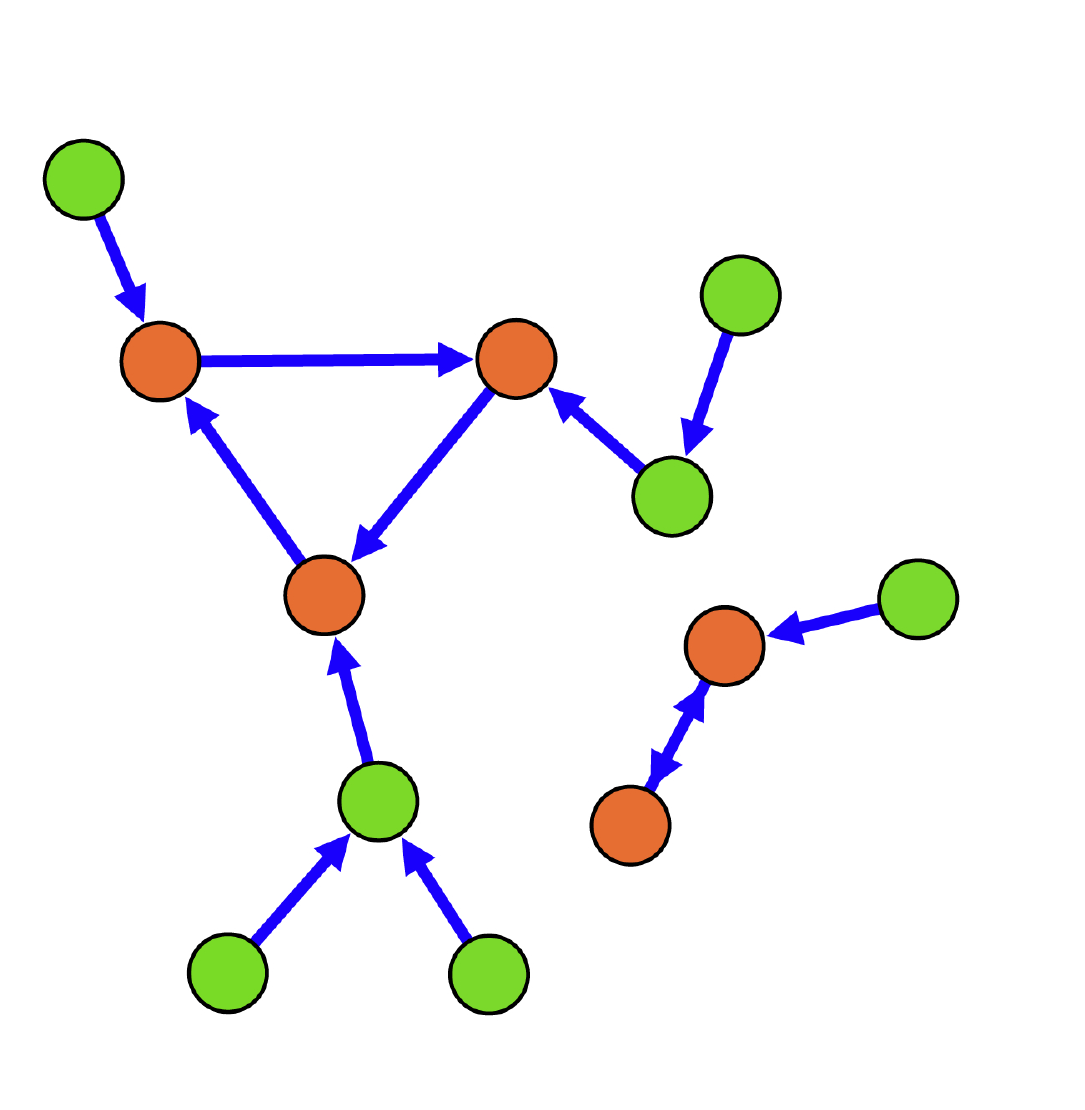}\label{fig:maximal_correction_input}}\hspace{1cm}
 \subfloat[Branch switch]{\includegraphics[width=0.5\columnwidth]{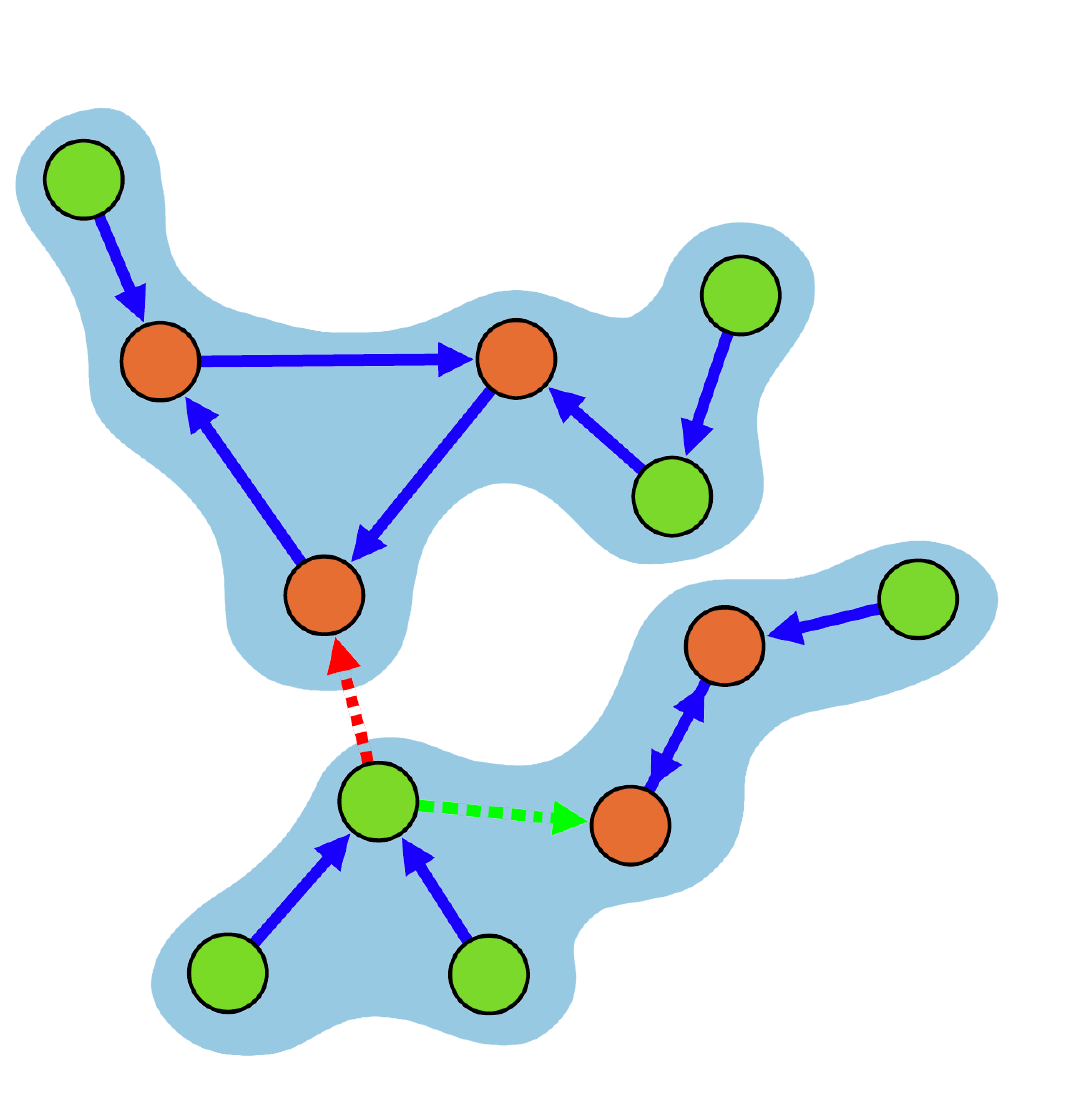}\label{fig:maximal_correction_branch}}\hspace{1cm}
 \subfloat[SCC switch]{\includegraphics[width=0.5\columnwidth]{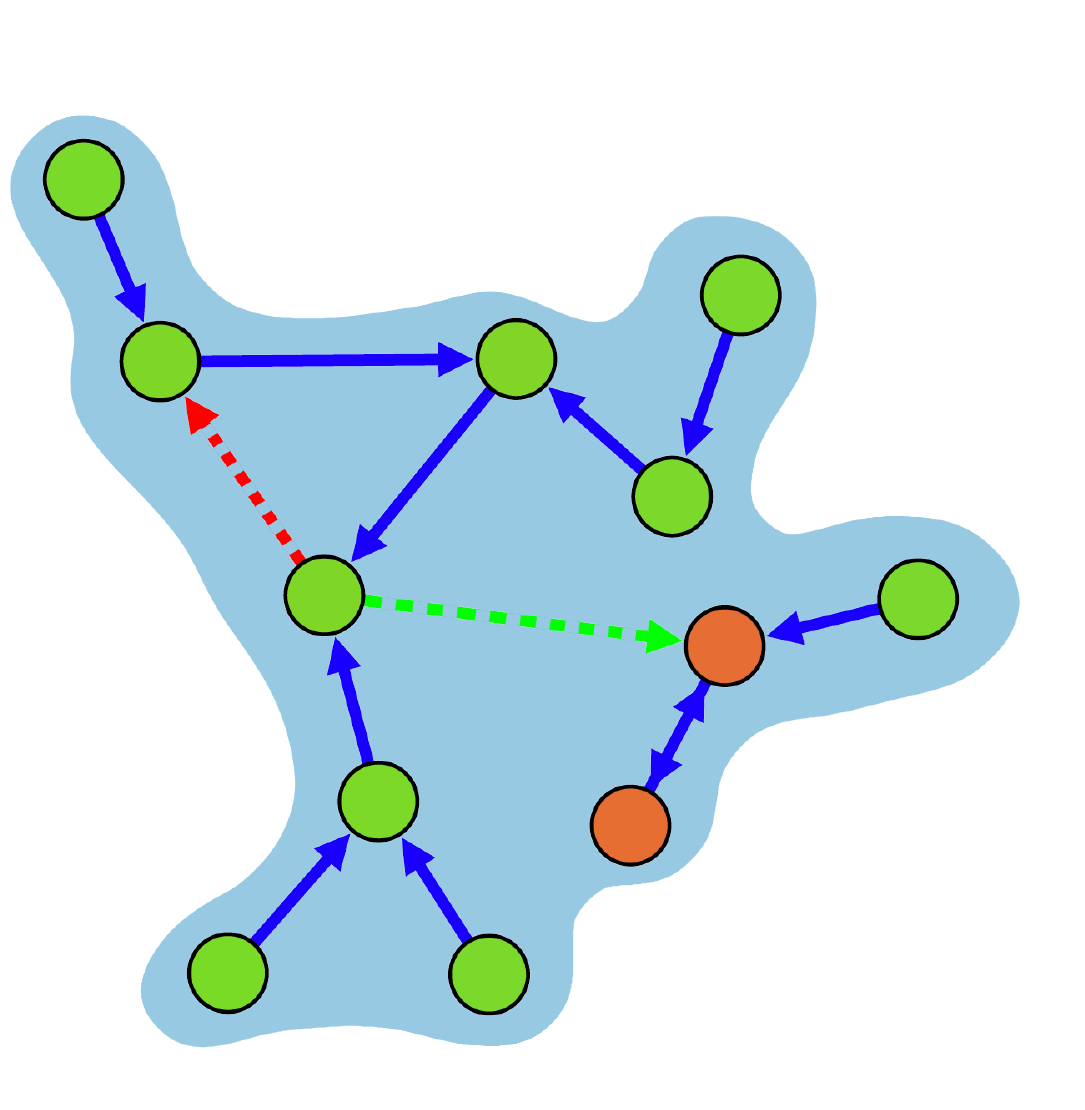}\label{fig:maximal_correction_scc}}
   \caption{Maximal correction step applied to an input community (a) containing a node with non maximal local gain. If the node is in one of the branches (b), the switch is applied to its entire tail. If the node is in the SCC (c), its tail spans the complete community which leads to merging that entire community with the destination of the switch.}
   \label{fig:maximal_correction}
 \end{figure*}
 
When no more correction is required to ensure a positive contribution of every node in the network, we try to further optimize the cost function by allowing nodes to switch from one community to another (function \emph{Switch}, Algorithm \ref{algo_sync_louvain}, step \ref{algo_maximal_corr}). While the initial assignment was optimal if all nodes were alone, this is not the case anymore because communities evolved from singletons to more complex structures spanned by the assignment graph. Based on the current distribution, one can compute the gain of switching each individual node to one of its neighbouring communities, \eg $$\Delta Q_M(i\rightarrow c_j) = Q_M(i\rightarrow c_j) - Q_M(i\rightarrow c_i).$$ However, since nodes belong to the assignment tree-like structure, when a node from a branch switches its community assignment, it will force the other nodes within its tail to follow as represented in Figure \ref{fig:maximal_correction_branch}. If the vertex was in the SCC of the community before the switch, it will force the entire community to follow, see Figure \ref{fig:maximal_correction_scc}. So, for every  node with a positive switching gain, $\Delta Q_M(i\rightarrow c_j) > 0$, our algorithm computes the gain of switching the community of its entire tail and only accepts moves that produce positive global gains. We refer to this procedure as the maximal correction step.
If we consider the modularity as the cost function to optimize, then the assignment switch of node $i$ with a positive switching gain and a tail $T$ (the set of nodes having a directed path leading to $i$ in the assignment graph), to the community $c$ will be accepted only if $$\sum\limits_{k\in T\cup\{i\}}\Delta Q_M(k\rightarrow c) + \sum\limits_{k\in T\cup \{i\}} Q_M(k\rightarrow T\cup\{i\}) > 0.$$
When a node switch is accepted, the destination of the switch is computed as the best single node assignment within the new community.

It is worth noticing that while the positive correction steps only increase the number of communities, the maximal correction steps will in general decrease the number of communities by reassigning nodes that were in a SCC (effectively breaking the SCC), \eg Figure \ref{fig:maximal_correction_scc}.

The maximal correction step does not use any other information than the current community distribution and the community distribution is updated after all the assignment switches have been computed. This allows an efficient use of a parallel processor architecture where each core can handle a set of nodes independently.

This sequence of positive corrections followed by maximal corrections will be repeated until all the nodes are assigned to the community with a maximal non-negative gain in the current distribution. However, the edge weight distribution might prevent this iterative scheme to converge. Indeed, situations where mutually attractive or repulsive nodes permanently switch their community assignments often happen, producing an infinite sequence of positive/maximal corrections. This kind of phenomenon also arises on consensus problem for synchronous multi-agent systems \cite{sarlette2008global,carli2010gossip}. While various strategies can be applied, our algorithm is designed to avoid this obstacle by accepting each individual maximal correction with a probability $p<1$. This probability somehow balances the quality of the clustering with the computational time required to meet a stable state and guarantees the convergence. Indeed, when $p<1$, one can consider each possible assignment graph as a state in a discrete-time Markov chain. One can then show that for any initial state, there exists at least one path leading to a steady state (\ie an assignment graph without any strictly positive switch), which in turn guarantees the convergence of the algorithm with probability $1$. When $p$ is large, more corrections are accepted increasing the overall quality of the communities but also the number of iteration to reach convergence. In all the experiments conducted in this paper, $p$ was set to $0.8$.

When no more corrections provide a strictly positive gain, the community graph is collapsed as in the Louvain method (Algorithm \ref{algo_sync_louvain}, step \ref{algo_maximal_collapse}). Then, this aggregated graph is used as a new input for the procedure to provide a new hierarchical level of clustering. The algorithm stops when there is no community structure in the last collapsed graph and all meta nodes remain single communities.

In the next section we apply our algorithm to benchmark networks and compare it to other popular methods. We show that our algorithm qualitatively performs almost as well as the Louvain method and outperforms other methods while achieving a very small time complexity.

\section{Benchmark Performance}
\begin{figure*}
 \includegraphics[width=1.99\columnwidth]{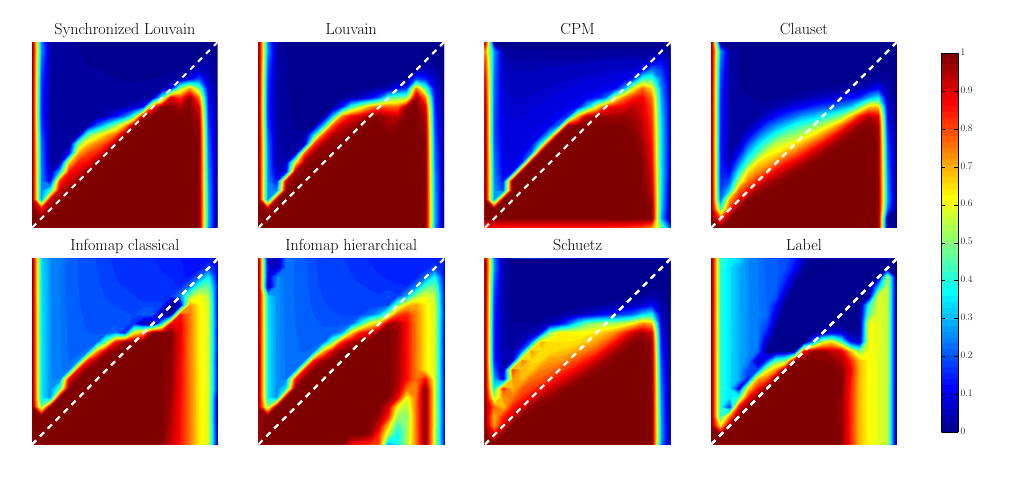}%
 \caption{Interpolation of the average value of NMI for each pair of mixing parameter $(\mu_T, \mu_W)$. The $x-$axis captures the value of $\mu_T$ and the $y-$axis captures the value of $\mu_W$. The mean value of NMI is computed for each of the $7$ algorithms (CPM stands for the Louvain method applied to a different cost function than modularity) on $50$ realisations per mixing parameter pair of the LFR benchmark with $10000$ nodes, average degree of $30$ and community size in the range $[100,1000]$.}
 \label{fig:bench_results_10000}
\end{figure*}
We compare our algorithm with $6$ other popular community detection algorithms: the Louvain method proposed by Blondel \etal \cite{Blondel2008} that inspired our algorithm,  the fast modularity proposed by Clauset, Newman \& Moore \cite{Clauset2004}, the Infomap method in its classical and hierarchical versions, developed by Rosvall \& Bergstrom \cite{Rosvall2008,Rosvall2011}, the multi-step greedy and vertex mover algorithm by Schuetz \& Caflisch \cite{Schuetz2008} and finally the label propagation method introduced by Raghavan \etal~\cite{Raghavan2007} and implemented in the Igraph package \cite{Csardi2006}.

The algorithms were compared on the popular LFR benchmark \cite{Lancichinetti2009} that produces weighted and directed networks controlled by several parameters such as the size of the network, the degree distribution or the community index of each node. The benchmark graphs created have a \enquote{ground truth} community distribution defined a priori that can be used to infer the quality of clustering algorithms.
Once the degree, the strength and the community index of each node have been defined, weighted directed edges are added to the graph to match as closely as possible the previously chosen degree distribution, while maintaining on average the value of two mixing parameters that induce the strength of the previously defined communities. The mixing parameter on the topology $\mu_T$ defines the expected proportion of edges for each node that is directed outside its community. The mixing parameter on the weights $\mu_W$ defines the expected proportion of the total weight of each node that is directed outside of its community. In other words, $\mu_T$ influences the internal and external degree of each node while $\mu_W$ controls the internal and external strength, \ie the weighted degree. Large values of mixing produce networks with either more edges or edges with higher weight outside of the a priori defined communities, effectively reducing the strength of this structure and making it harder to extract.

To compare the extracted partitions of each algorithm with the a priori known community structure of the benchmark graphs, we used the normalized mutual information (NMI) \cite{Danon2005}. The NMI is a similarity measure between two partitions $X$ and $Y$ that represents their normalized mutual entropy and can be computed as 
\begin{equation*}
NMI(X,Y) = \frac{2\mathcal{I}\left(X,Y\right)}{\mathcal{H}\left(X\right)+\mathcal{H}\left(Y\right)},
\label{eq:nmi}
\end{equation*}
where $\mathcal{H}\left(X\right)$ is the entropy of the partition $X$ and $\mathcal{I}\left(X,Y\right)$ is the mutual information of the partitions $X$ and $Y$, given respectively by
\begin{eqnarray*}
\mathcal{H}\left(X\right) &=& \sum\limits_r \frac{n_r}{N}\log\frac{n_r}{N},\\
\mathcal{I}\left(X,Y\right) &=& \sum\limits_{r,\,s} \frac{n_{rs}}{N}\log N\frac{n_r\,n_s}{n_{rs}},
\label{eq:entropy}
\end{eqnarray*}
with $n_r$ the number of nodes in community $r$ and $n_{rs}$ the number of common nodes in community $r$ of distribution $X$ and community $s$ of distribution $Y$. The NMI ranges in $[0,1]$ and equals $1$ only for identical distributions.

The results of our experiments on graphs with $10.000$ nodes are presented in Figure \ref{fig:bench_results_10000}. Each panel represents the average result of an algorithm or a specific cost function. The $x$-axis indicates the value of $\mu_T$ and the $y$-axis captures the value of $\mu_W$. Each figure displays a smooth interpolation of the average value of NMI computed on $50$ graphs for each couple of mixing parameters $(\mu_T, \mu_W) \in [0,1]\times[0,1]$ discretized with a step size of $0.05$.
The other parameters of the benchmark graphs were set to an average degree of $30$ and a community size distribution in the range $\left[100,1000\right]$. Very similar results have been obtained with different sets of parameters. 
  
The top left panel shows the results of our algorithm applied to the modularity cost function and shows that its performance is similar to the classical Louvain method (second panel of first row). Indeed, the area of the region covered with large values of NMI are almost identical and while the values of NMI are slightly higher for the original Louvain method, it also suffers of a steeper decay of NMI for large values of $\mu_T$. 

Both algorithms clearly outperform the algorithm of Clauset ($4^{th}$ panel, first row) and the Label propagation method ($4^{th}$ panel, second row). The CPM cost function provides good partitions but, its average performance is not superior than the modularity. However, the $\gamma$ parameter was not optimized to provide the best possible solution which would have increased the overall quality. The performance of the CPM is similar when optimized using our algorithm or the Louvain method, so only one panel is represented in Figure \ref{fig:bench_results_10000}.

All algorithms perform badly above the dashed white diagonal. This diagonal separation of the experimental plane represents the set of points where $\mu_T = \mu_W$. The LFR benchmark produces networks where the expected weight of an internal and an external edge can be computed as 
\begin{equation}
\left\langle w_{int}\right\rangle = \frac{1-\mu_W}{1-\mu_T}k_i\, ,\quad \left\langle w_{ext}\right\rangle = \frac{\mu_W}{\mu_T}k_i,
\label{eq:int_ext_weight}
\end{equation}
where $k_i$  is the strength of node $i$. This implies that the white dashed diagonal represents the set of points where on average, the expected weight of an edge is identical whether the edge is internal or external. This implies that above the diagonal, edge weights start to be higher outside of the communities, which makes those communities harder to extract and explains why all algorithms suffer in this range. However, it is questionable if the a-priori defined communities are still relevant in this range and if it remains desirable that the algorithms extract them. Clearly, the algorithm of Schuetz and Caflisch has slightly better performance than all other algorithms in this range of values, but this method produces values of NMI that are lower everywhere else in the experimental plane.

Finally the Infomap of Rosvall and Bergstrom performs correctly in its classical version but does not produce good hierarchical modules. The performance decay for large values of $\mu_T$ is smoother, however it starts at a smaller parameter value (around $0.6$) producing a clear band where our algorithm outperforms this method.

\begin{figure}
	\includegraphics[width=0.99\columnwidth]{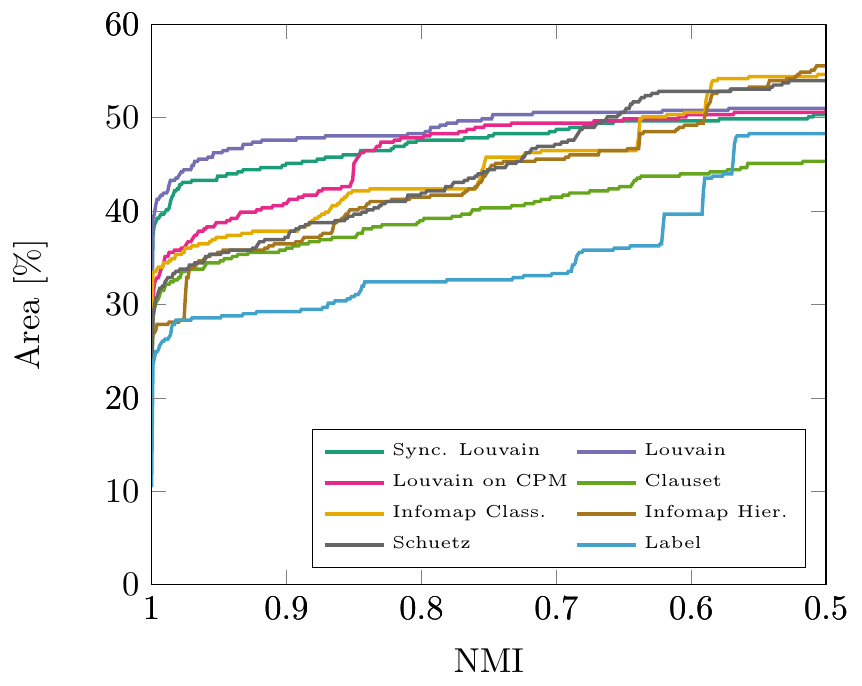}%
	\caption{Cumulative area covered in the experimental planed of Figure \ref{fig:bench_results_10000} by each algorithm tested for decreasing value of NMI.}
	\label{fig:bench_quality}
\end{figure}

A comparison between the different algorithms is presented in Figure \ref{fig:bench_quality} where each curve corresponds to the cumulative area covered by one algorithm for decreasing value of NMI. The values of NMI were limited to $0.5$ since lower values are irrelevant, the quality of the extracted communities being as bad as communities created at random. One can see that the Louvain method clearly dominates all other algorithms with almost $50\%$ of mixing parameter couple for all NMI threshold larger than $0.7$. However, the distance between our algorithm and the Louvain method is rather small for any NMI threshold. The only other method that produces good results is CPM which is the Louvain method applied to the constant Potts model cost function. As mentioned previously, both the Infomap and the algorithm by Schuetz and Caflisch have a smaller area of large NMI. However, those methods have a smoother decay for increasing NMI threshold, leading to a larger region of mixing parameter for NMI values smaller than $0.65$. However, depending on the application, such values of NMI might already be considered too small to produce relevant community structures. Finally, both the algorithm of Clauset \etal and the Label propagation method clearly have worse performance than the other methods when applied to weighted directed networks.

\begin{figure*}
 \includegraphics[width=0.99\columnwidth]{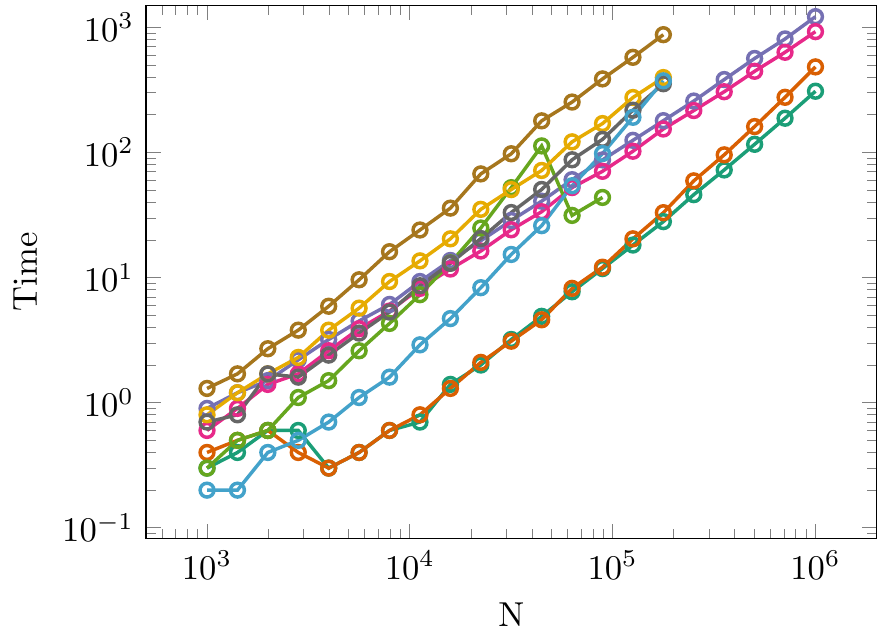}%
 \includegraphics[width=0.95\columnwidth]{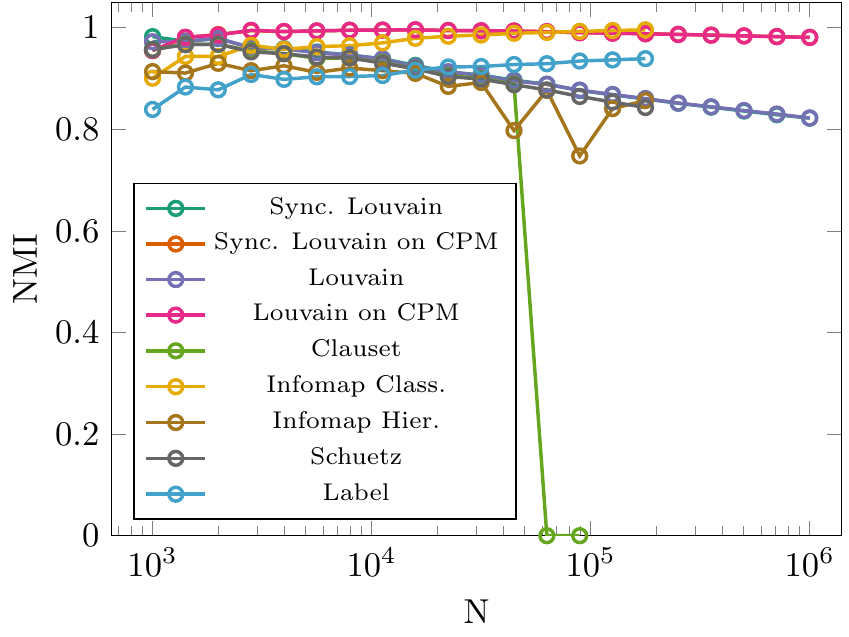}%
 \caption{Average computational time and NMI for each tested algorithm applied to networks of size $10^3$ up to $10^6$ nodes. The average degree was set to $N/100$ and community size are in $[N/100, N/10]$. Each circle corresponds to the average computational time for several mixing parameter couple $(\mu_T, \mu_W)$ with $25$ graph realisations per couple.}
 \label{fig:bench_results_time}
 \end{figure*}

This shows that the qualitative performance of our algorithm is similar to other popular techniques in community detection. However, the main achievement of our algorithm is its very small time complexity together with its highly parallelizable behaviour.
We analysed the computational time required by each algorithm to extract community partitions for benchmark graphs growing from $10^3$ to $10^6$ nodes. The average degree was set to one hundredth of the number of nodes in the graph and the community sizes were chosen between one hundredth and one tenth of the number of nodes. The results are presented in Figure  \ref{fig:bench_results_time}. The left panel shows the average computational time for each method and the right panel shows the average NMI value. Each algorithm was tested against a set of mixing parameter couples with $\mu_T,\mu_W \in \{0.1,0.3,0.5,0.7,0.9\}$, $\mu_W \leq \mu_T$.

\begin{figure}
\includegraphics[width=0.95\columnwidth]{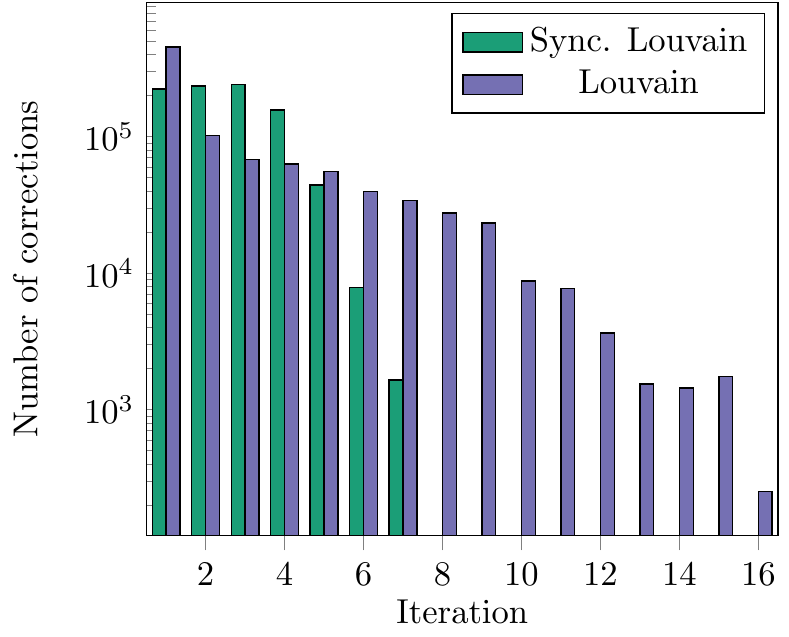}
 \caption{Typical number of corrections per iteration for our synchronized algorithm and the classical Louvain method.}
 \label{fig:number_correction}
 \end{figure}
 
To have a fair comparison, all the different algorithms were tested with the \texttt{C++} implementation provided by the authors (except for the label propagation algorithm for which we used the implementation of the Igraph $c$ package) and executed on dedicated single processors. This implies that no external tasks could have slowed the processor and perturbed the results.
Some algorithms were not tested for large graph sizes when the computational time required was larger than $1500$ seconds and the method by Clauset \etal was discarded early because of memory issues leading to poor quality for the communities. It is clear from Figure \ref{fig:bench_results_time} that our algorithm outperforms all the other methods. We observed that our synchronized algorithm performs between $5$ and $20$ times faster than the classical Louvain method using modularity. Even better performance could have been reached using a dedicated parallel processor architecture. However, our algorithm is less suitable to analyse unweighted networks. We observe that when applied to unweighted networks, our algorithm slows down due to an increased number of corrections required. When the assignment graph is created, each assignment is solely controlled by the degree of the neighbours for the modularity or is performed at random for the constant Potts model. Consequently, the number of corrections to extract an assignment graph revealing the community structure of the graph increases significantly.

One can observe that even if our algorithm is much faster, the quality of our method and the Louvain method are so similar that the NMI curves coalesce. The decay of quality for large sizes is due to the well known resolution limit problem of modularity previously mentioned.  This observation is confirmed by the $2$ curves for the algorithms applied to the constant Potts model (CPM) \cite{Traag2011} which leads to better quality of the partitions.
 
The explanation behind the large speed-up of our algorithm lies in two observations. First, the two types of correction we introduced allow to change the community assignment of a large number of nodes at the same time by considering the assignment graph and the tail of each node. This leads to a clear computational gain compared to sequential modifications. Another observation is that even if the total number of corrections is about the same for both algorithms, our method allows to find a steady partition of the nodes in less iterations as displayed in Figure \ref{fig:number_correction}. Each iteration requires to loop over all the nodes and edges of the graph to check for possible increases in the cost function. By reducing the required number of iterations to reach convergence, our algorithm extracts much faster the first hierarchical level of communities which is the dominant factor in the total time complexity.

\section{Conclusion}

Community detection is a challenging task but it provides powerful insights about the complex structures of large networks and allows to analyse complex phenomena at different scales. In this paper, we demonstrate how the Louvain method can be altered to provide a highly synchronizable algorithm to extract community structures.
Extensive testing has been done on the popular LFR benchmark and the partitions produced by our algorithm are of quality similar to the original method and even outperforms other popular algorithms. In the meantime, our method exhibits a much smaller computational time by allowing multiple nodes to switch their community assignment synchronously and reaching a stable partition with less iterations.
In our future works, we plan to provide a parallel implementation of the software and to analyse weighted networks of unprecedented sizes.

\begin{acknowledgments}
\noindent The authors would like to thank  V. Traag and J. Hendrickx for inspiring discussions.

\noindent This paper presents research results supported by the Concerted Research Action (ARC) \enquote{Large Graphs and Networks} of the French Community of Belgium and the Belgian Network DYSCO (Dynamical Systems, Control, and Optimization), funded by the Interuniversity Attraction Poles Programme,
initiated by the Belgian State, Science Policy Office.

\noindent Computational resources have been provided by the supercomputing facilities of the Universit\'e catholique de Louvain (CISM/UCL) and the Consortium des \'Equipements de Calcul Intensif en F\'ed\'eration Wallonie Bruxelles (C\'ECI) funded by the Fond de la Recherche Scientifique de Belgique (F.R.S.-FNRS).
\end{acknowledgments}

\bibliography{bibliography_mendeley}

\begin{thebibliography}{44}%
\makeatletter
\providecommand \@ifxundefined [1]{%
 \@ifx{#1\undefined}
}%
\providecommand \@ifnum [1]{%
 \ifnum #1\expandafter \@firstoftwo
 \else \expandafter \@secondoftwo
 \fi
}%
\providecommand \@ifx [1]{%
 \ifx #1\expandafter \@firstoftwo
 \else \expandafter \@secondoftwo
 \fi
}%
\providecommand \natexlab [1]{#1}%
\providecommand \enquote  [1]{``#1''}%
\providecommand \bibnamefont  [1]{#1}%
\providecommand \bibfnamefont [1]{#1}%
\providecommand \citenamefont [1]{#1}%
\providecommand \href@noop [0]{\@secondoftwo}%
\providecommand \href [0]{\begingroup \@sanitize@url \@href}%
\providecommand \@href[1]{\@@startlink{#1}\@@href}%
\providecommand \@@href[1]{\endgroup#1\@@endlink}%
\providecommand \@sanitize@url [0]{\catcode `\\12\catcode `\$12\catcode
  `\&12\catcode `\#12\catcode `\^12\catcode `\_12\catcode `\%12\relax}%
\providecommand \@@startlink[1]{}%
\providecommand \@@endlink[0]{}%
\providecommand \url  [0]{\begingroup\@sanitize@url \@url }%
\providecommand \@url [1]{\endgroup\@href {#1}{\urlprefix }}%
\providecommand \urlprefix  [0]{URL }%
\providecommand \Eprint [0]{\href }%
\providecommand \doibase [0]{http://dx.doi.org/}%
\providecommand \selectlanguage [0]{\@gobble}%
\providecommand \bibinfo  [0]{\@secondoftwo}%
\providecommand \bibfield  [0]{\@secondoftwo}%
\providecommand \translation [1]{[#1]}%
\providecommand \BibitemOpen [0]{}%
\providecommand \bibitemStop [0]{}%
\providecommand \bibitemNoStop [0]{.\EOS\space}%
\providecommand \EOS [0]{\spacefactor3000\relax}%
\providecommand \BibitemShut  [1]{\csname bibitem#1\endcsname}%
\let\auto@bib@innerbib\@empty
\bibitem [{\citenamefont {Halevy}\ \emph {et~al.}(2009)\citenamefont {Halevy},
  \citenamefont {Norvig},\ and\ \citenamefont {Pereira}}]{halevy2009}%
  \BibitemOpen
  \bibfield  {author} {\bibinfo {author} {\bibfnamefont {A.}~\bibnamefont
  {Halevy}}, \bibinfo {author} {\bibfnamefont {P.}~\bibnamefont {Norvig}}, \
  and\ \bibinfo {author} {\bibfnamefont {F.}~\bibnamefont {Pereira}},\ }\href
  {\doibase 10.1109/MIS.2009.36} {\bibfield  {journal} {\bibinfo  {journal}
  {IEEE Intelligent Systems}\ }\textbf {\bibinfo {volume} {24}},\ \bibinfo
  {pages} {8} (\bibinfo {year} {2009})}\BibitemShut {NoStop}%
\bibitem [{\citenamefont {Torralba}\ \emph {et~al.}(2008)\citenamefont
  {Torralba}, \citenamefont {Fergus},\ and\ \citenamefont
  {Freeman}}]{torralba2008}%
  \BibitemOpen
  \bibfield  {author} {\bibinfo {author} {\bibfnamefont {A.}~\bibnamefont
  {Torralba}}, \bibinfo {author} {\bibfnamefont {R.}~\bibnamefont {Fergus}}, \
  and\ \bibinfo {author} {\bibfnamefont {W.~T.}\ \bibnamefont {Freeman}},\
  }\href {http://www.ncbi.nlm.nih.gov/pubmed/18787244} {\bibfield  {journal}
  {\bibinfo  {journal} {IEEE Transactions on Pattern Analysis and Machine
  Intelligence}\ }\textbf {\bibinfo {volume} {30}},\ \bibinfo {pages} {1958}
  (\bibinfo {year} {2008})}\BibitemShut {NoStop}%
\bibitem [{\citenamefont {Skillicorn}\ and\ \citenamefont
  {Talia}(2012)}]{skillicorn2012}%
  \BibitemOpen
  \bibfield  {author} {\bibinfo {author} {\bibfnamefont {D.}~\bibnamefont
  {Skillicorn}}\ and\ \bibinfo {author} {\bibfnamefont {D.}~\bibnamefont
  {Talia}},\ }\href
  {http://www.cai.sk/ojs/index.php/cai/article/viewArticle/488} {\bibfield
  {journal} {\bibinfo  {journal} {Computing and Informatics}\ }\textbf
  {\bibinfo {volume} {21}},\ \bibinfo {pages} {347} (\bibinfo {year}
  {2012})}\BibitemShut {NoStop}%
\bibitem [{\citenamefont {Albert}\ and\ \citenamefont
  {Barab\'{a}si}(2007)}]{Albert2007}%
  \BibitemOpen
  \bibfield  {author} {\bibinfo {author} {\bibfnamefont {R.}~\bibnamefont
  {Albert}}\ and\ \bibinfo {author} {\bibfnamefont {A.-L.}\ \bibnamefont
  {Barab\'{a}si}},\ }\href {http://arxiv.org/abs/cond-mat/0106096} {\bibfield
  {journal} {\bibinfo  {journal} {World Wide Web Internet And Web Information
  Systems}\ }\bibinfo {series} {Lecture Notes In Physics},\ \textbf {\bibinfo
  {volume} {74}},\ \bibinfo {pages} {78} (\bibinfo {year} {2007})}\BibitemShut
  {NoStop}%
\bibitem [{\citenamefont {Lazer}\ \emph {et~al.}(2009)\citenamefont {Lazer},
  \citenamefont {Pentland}, \citenamefont {Adamic}, \citenamefont {Aral},
  \citenamefont {Barabasi}, \citenamefont {Brewer}, \citenamefont {Christakis},
  \citenamefont {Contractor}, \citenamefont {Fowler}, \citenamefont {Gutmann},
  \citenamefont {Jebara}, \citenamefont {King}, \citenamefont {Macy},
  \citenamefont {Roy},\ and\ \citenamefont {{Van Alstyne}}}]{lazer2009}%
  \BibitemOpen
  \bibfield  {author} {\bibinfo {author} {\bibfnamefont {D.}~\bibnamefont
  {Lazer}}, \bibinfo {author} {\bibfnamefont {A.}~\bibnamefont {Pentland}},
  \bibinfo {author} {\bibfnamefont {L.}~\bibnamefont {Adamic}}, \bibinfo
  {author} {\bibfnamefont {S.}~\bibnamefont {Aral}}, \bibinfo {author}
  {\bibfnamefont {A.-L.}\ \bibnamefont {Barabasi}}, \bibinfo {author}
  {\bibfnamefont {D.}~\bibnamefont {Brewer}}, \bibinfo {author} {\bibfnamefont
  {N.}~\bibnamefont {Christakis}}, \bibinfo {author} {\bibfnamefont
  {N.}~\bibnamefont {Contractor}}, \bibinfo {author} {\bibfnamefont
  {J.}~\bibnamefont {Fowler}}, \bibinfo {author} {\bibfnamefont
  {M.}~\bibnamefont {Gutmann}}, \bibinfo {author} {\bibfnamefont
  {T.}~\bibnamefont {Jebara}}, \bibinfo {author} {\bibfnamefont
  {G.}~\bibnamefont {King}}, \bibinfo {author} {\bibfnamefont {M.}~\bibnamefont
  {Macy}}, \bibinfo {author} {\bibfnamefont {D.}~\bibnamefont {Roy}}, \ and\
  \bibinfo {author} {\bibfnamefont {M.}~\bibnamefont {{Van Alstyne}}},\ }\href
  {\doibase 10.1126/science.1167742} {\bibfield  {journal} {\bibinfo  {journal}
  {Science}\ }\textbf {\bibinfo {volume} {323}},\ \bibinfo {pages} {721}
  (\bibinfo {year} {2009})}\BibitemShut {NoStop}%
\bibitem [{\citenamefont {Onnela}\ \emph {et~al.}(2007)\citenamefont {Onnela},
  \citenamefont {Saram\"{a}ki}, \citenamefont {Hyv\"{o}nen}, \citenamefont
  {Szab\'{o}}, \citenamefont {Lazer}, \citenamefont {Kaski}, \citenamefont
  {Kert\'{e}sz},\ and\ \citenamefont {Barab\'{a}si}}]{Onnela2007}%
  \BibitemOpen
  \bibfield  {author} {\bibinfo {author} {\bibfnamefont {J.~P.}\ \bibnamefont
  {Onnela}}, \bibinfo {author} {\bibfnamefont {J.}~\bibnamefont
  {Saram\"{a}ki}}, \bibinfo {author} {\bibfnamefont {J.}~\bibnamefont
  {Hyv\"{o}nen}}, \bibinfo {author} {\bibfnamefont {G.}~\bibnamefont
  {Szab\'{o}}}, \bibinfo {author} {\bibfnamefont {D.}~\bibnamefont {Lazer}},
  \bibinfo {author} {\bibfnamefont {K.}~\bibnamefont {Kaski}}, \bibinfo
  {author} {\bibfnamefont {J.}~\bibnamefont {Kert\'{e}sz}}, \ and\ \bibinfo
  {author} {\bibfnamefont {A.~L.}\ \bibnamefont {Barab\'{a}si}},\ }\href
  {http://www.pubmedcentral.nih.gov/articlerender.fcgi?artid=1863470\&tool=pmcentrez\&rendertype=abstract}
  {\bibfield  {journal} {\bibinfo  {journal} {Proceedings of the National
  Academy of Sciences of the United States of America}\ }\textbf {\bibinfo
  {volume} {104}},\ \bibinfo {pages} {7332} (\bibinfo {year}
  {2007})}\BibitemShut {NoStop}%
\bibitem [{\citenamefont {Kumpula}\ \emph {et~al.}(2009)\citenamefont
  {Kumpula}, \citenamefont {Onnela}, \citenamefont {Saram\"{a}ki},
  \citenamefont {Kert\'{e}sz},\ and\ \citenamefont {Kaski}}]{Kumpula2009}%
  \BibitemOpen
  \bibfield  {author} {\bibinfo {author} {\bibfnamefont {J.}~\bibnamefont
  {Kumpula}}, \bibinfo {author} {\bibfnamefont {J.-P.}\ \bibnamefont {Onnela}},
  \bibinfo {author} {\bibfnamefont {J.}~\bibnamefont {Saram\"{a}ki}}, \bibinfo
  {author} {\bibfnamefont {J.}~\bibnamefont {Kert\'{e}sz}}, \ and\ \bibinfo
  {author} {\bibfnamefont {K.}~\bibnamefont {Kaski}},\ }\href {\doibase
  10.1016/j.cpc.2008.12.016} {\bibfield  {journal} {\bibinfo  {journal}
  {Computer Physics Communications}\ }\textbf {\bibinfo {volume} {180}},\
  \bibinfo {pages} {517} (\bibinfo {year} {2009})}\BibitemShut {NoStop}%
\bibitem [{\citenamefont {Albert}\ \emph {et~al.}(1999)\citenamefont {Albert},
  \citenamefont {Jeong},\ and\ \citenamefont {Barabasi}}]{Albert1999}%
  \BibitemOpen
  \bibfield  {author} {\bibinfo {author} {\bibfnamefont {R.}~\bibnamefont
  {Albert}}, \bibinfo {author} {\bibfnamefont {H.}~\bibnamefont {Jeong}}, \
  and\ \bibinfo {author} {\bibfnamefont {A.-L.}\ \bibnamefont {Barabasi}},\
  }\href {http://arxiv.org/abs/cond-mat/9907038} {\bibfield  {journal}
  {\bibinfo  {journal} {Nature}\ }\textbf {\bibinfo {volume} {401}},\ \bibinfo
  {pages} {5} (\bibinfo {year} {1999})}\BibitemShut {NoStop}%
\bibitem [{\citenamefont {Newman}(2000)}]{Newman2000}%
  \BibitemOpen
  \bibfield  {author} {\bibinfo {author} {\bibfnamefont {M.~E.~J.}\
  \bibnamefont {Newman}},\ }\href {http://arxiv.org/abs/cond-mat/0007214}
  {\bibfield  {journal} {\bibinfo  {journal} {Proceedings of the National
  Academy of Sciences of the United States of America}\ }\textbf {\bibinfo
  {volume} {98}},\ \bibinfo {pages} {7} (\bibinfo {year} {2000})}\BibitemShut
  {NoStop}%
\bibitem [{\citenamefont {Shaw}\ and\ \citenamefont {Tunc}(2012)}]{Tunk2012}%
  \BibitemOpen
  \bibfield  {author} {\bibinfo {author} {\bibfnamefont {L.}~\bibnamefont
  {Shaw}}\ and\ \bibinfo {author} {\bibfnamefont {I.}~\bibnamefont {Tunc}},\
  }in\ \href {\doibase 10.1007/978-3-642-32615-8_49} {\emph {\bibinfo
  {booktitle} {Bio-Inspired Models of Network, Information, and Computing
  Systems}}},\ Vol.~\bibinfo {volume} {87}\ (\bibinfo {year} {2012})\ pp.\
  \bibinfo {pages} {519--520}\BibitemShut {NoStop}%
\bibitem [{\citenamefont {Wu}\ and\ \citenamefont {Liu}(2008)}]{Wu2008}%
  \BibitemOpen
  \bibfield  {author} {\bibinfo {author} {\bibfnamefont {X.}~\bibnamefont
  {Wu}}\ and\ \bibinfo {author} {\bibfnamefont {Z.}~\bibnamefont {Liu}},\
  }\href {\doibase 10.1016/j.physa.2007.09.039} {\bibfield  {journal} {\bibinfo
   {journal} {Physica A: Statistical Mechanics and its Applications}\ }\textbf
  {\bibinfo {volume} {387}},\ \bibinfo {pages} {623} (\bibinfo {year}
  {2008})}\BibitemShut {NoStop}%
\bibitem [{\citenamefont {Liu}\ and\ \citenamefont {Hu}(2005)}]{Liu2005}%
  \BibitemOpen
  \bibfield  {author} {\bibinfo {author} {\bibfnamefont {Z.}~\bibnamefont
  {Liu}}\ and\ \bibinfo {author} {\bibfnamefont {B.}~\bibnamefont {Hu}},\
  }\href {\doibase 10.1209/epl/i2004-10550-5} {\bibfield  {journal} {\bibinfo
  {journal} {Europhysics Letters}\ }\textbf {\bibinfo {volume} {72}},\ \bibinfo
  {pages} {315} (\bibinfo {year} {2005})}\BibitemShut {NoStop}%
\bibitem [{\citenamefont {Guimer\`{a}}\ \emph {et~al.}(2005)\citenamefont
  {Guimer\`{a}}, \citenamefont {Mossa}, \citenamefont {Turtschi},\ and\
  \citenamefont {Amaral}}]{Guimera2005a}%
  \BibitemOpen
  \bibfield  {author} {\bibinfo {author} {\bibfnamefont {R.}~\bibnamefont
  {Guimer\`{a}}}, \bibinfo {author} {\bibfnamefont {S.}~\bibnamefont {Mossa}},
  \bibinfo {author} {\bibfnamefont {A.}~\bibnamefont {Turtschi}}, \ and\
  \bibinfo {author} {\bibfnamefont {L.~A.~N.}\ \bibnamefont {Amaral}},\ }\href
  {http://arxiv.org/abs/cond-mat/0312535} {\bibfield  {journal} {\bibinfo
  {journal} {Proceedings of the National Academy of Sciences of the United
  States of America}\ }\textbf {\bibinfo {volume} {102}},\ \bibinfo {pages}
  {7794} (\bibinfo {year} {2005})}\BibitemShut {NoStop}%
\bibitem [{\citenamefont {Guimera}\ and\ \citenamefont
  {Amaral}(2005)}]{Guimera2005}%
  \BibitemOpen
  \bibfield  {author} {\bibinfo {author} {\bibfnamefont {R.}~\bibnamefont
  {Guimera}}\ and\ \bibinfo {author} {\bibfnamefont {L.~A.~N.}\ \bibnamefont
  {Amaral}},\ }\href {http://arxiv.org/abs/q-bio/0502035} {\bibfield  {journal}
  {\bibinfo  {journal} {Nature}\ }\textbf {\bibinfo {volume} {433}},\ \bibinfo
  {pages} {895} (\bibinfo {year} {2005})}\BibitemShut {NoStop}%
\bibitem [{\citenamefont {Browet}\ \emph {et~al.}(2011)\citenamefont {Browet},
  \citenamefont {Absil},\ and\ \citenamefont {{Van Dooren}}}]{Browet2011}%
  \BibitemOpen
  \bibfield  {author} {\bibinfo {author} {\bibfnamefont {A.}~\bibnamefont
  {Browet}}, \bibinfo {author} {\bibfnamefont {P.-A.}\ \bibnamefont {Absil}}, \
  and\ \bibinfo {author} {\bibfnamefont {P.}~\bibnamefont {{Van Dooren}}},\
  }\href {\doibase 10.1007/978-3-642-21073-0\_32} {\bibfield  {journal}
  {\bibinfo  {journal} {Combinatorial Image Analysis}\ }\bibinfo {series}
  {Combinatorial Image Analysis. Proceedings of the 14th International
  Workshop, IWCIA 2011},\ \bibinfo {pages} {358} (\bibinfo {year}
  {2011})}\BibitemShut {NoStop}%
\bibitem [{\citenamefont {Reichardt}\ and\ \citenamefont
  {Bornholdt}(2006)}]{Reichardt2006}%
  \BibitemOpen
  \bibfield  {author} {\bibinfo {author} {\bibfnamefont {J.}~\bibnamefont
  {Reichardt}}\ and\ \bibinfo {author} {\bibfnamefont {S.}~\bibnamefont
  {Bornholdt}},\ }\href {\doibase 10.1103/PhysRevE.74.016110} {\bibfield
  {journal} {\bibinfo  {journal} {Physical Review E}\ }\textbf {\bibinfo
  {volume} {74}},\ \bibinfo {pages} {16110} (\bibinfo {year}
  {2006})}\BibitemShut {NoStop}%
\bibitem [{\citenamefont {Newman}\ and\ \citenamefont
  {Girvan}(2004)}]{Newman2004}%
  \BibitemOpen
  \bibfield  {author} {\bibinfo {author} {\bibfnamefont {M.~E.~J.}\
  \bibnamefont {Newman}}\ and\ \bibinfo {author} {\bibfnamefont
  {M.}~\bibnamefont {Girvan}},\ }\href {http://arxiv.org/abs/cond-mat/0308217}
  {\bibfield  {journal} {\bibinfo  {journal} {Physical Review E - Statistical,
  Nonlinear and Soft Matter Physics}\ }\textbf {\bibinfo {volume} {69}},\
  \bibinfo {pages} {26113} (\bibinfo {year} {2004})}\BibitemShut {NoStop}%
\bibitem [{\citenamefont {Newman}(2004)}]{Newman2004a}%
  \BibitemOpen
  \bibfield  {author} {\bibinfo {author} {\bibfnamefont {M.~E.~J.}\
  \bibnamefont {Newman}},\ }\href {http://arxiv.org/abs/cond-mat/0407503}
  {\bibfield  {journal} {\bibinfo  {journal} {Physical Review E}\ }\textbf
  {\bibinfo {volume} {70}},\ \bibinfo {pages} {9} (\bibinfo {year}
  {2004})}\BibitemShut {NoStop}%
\bibitem [{\citenamefont {Leicht}\ and\ \citenamefont
  {Newman}(2007)}]{Leicht2007}%
  \BibitemOpen
  \bibfield  {author} {\bibinfo {author} {\bibfnamefont {E.~A.}\ \bibnamefont
  {Leicht}}\ and\ \bibinfo {author} {\bibfnamefont {M.~E.~J.}\ \bibnamefont
  {Newman}},\ }\href {http://arxiv.org/abs/0709.4500} {\bibfield  {journal}
  {\bibinfo  {journal} {Physical Review Letters}\ }\textbf {\bibinfo {volume}
  {100}},\ \bibinfo {pages} {118703} (\bibinfo {year} {2007})}\BibitemShut
  {NoStop}%
\bibitem [{\citenamefont {Kashtan}\ and\ \citenamefont
  {Alon}(2005)}]{Kashtan2005}%
  \BibitemOpen
  \bibfield  {author} {\bibinfo {author} {\bibfnamefont {N.}~\bibnamefont
  {Kashtan}}\ and\ \bibinfo {author} {\bibfnamefont {U.}~\bibnamefont {Alon}},\
  }\href {\doibase 10.1073/pnas.0503610102} {\bibfield  {journal} {\bibinfo
  {journal} {Proceedings of the National Academy of Sciences of the United
  States of America}\ }\textbf {\bibinfo {volume} {102}},\ \bibinfo {pages}
  {13773} (\bibinfo {year} {2005})}\BibitemShut {NoStop}%
\bibitem [{\citenamefont {Mucha}\ \emph {et~al.}(2010)\citenamefont {Mucha},
  \citenamefont {Richardson}, \citenamefont {Macon}, \citenamefont {Porter},\
  and\ \citenamefont {Onnela}}]{Mucha2010}%
  \BibitemOpen
  \bibfield  {author} {\bibinfo {author} {\bibfnamefont {P.~J.}\ \bibnamefont
  {Mucha}}, \bibinfo {author} {\bibfnamefont {T.}~\bibnamefont {Richardson}},
  \bibinfo {author} {\bibfnamefont {K.}~\bibnamefont {Macon}}, \bibinfo
  {author} {\bibfnamefont {M.~A.}\ \bibnamefont {Porter}}, \ and\ \bibinfo
  {author} {\bibfnamefont {J.-P.}\ \bibnamefont {Onnela}},\ }\href
  {http://arxiv.org/abs/0911.1824} {\bibfield  {journal} {\bibinfo  {journal}
  {Science}\ }\textbf {\bibinfo {volume} {328}},\ \bibinfo {pages} {31}
  (\bibinfo {year} {2010})}\BibitemShut {NoStop}%
\bibitem [{\citenamefont {Conover}\ \emph {et~al.}(2013)\citenamefont
  {Conover}, \citenamefont {Davis}, \citenamefont {Ferrara}, \citenamefont
  {McKelvey}, \citenamefont {Menczer},\ and\ \citenamefont
  {Flammini}}]{Conover2013}%
  \BibitemOpen
  \bibfield  {author} {\bibinfo {author} {\bibfnamefont {M.~D.}\ \bibnamefont
  {Conover}}, \bibinfo {author} {\bibfnamefont {C.}~\bibnamefont {Davis}},
  \bibinfo {author} {\bibfnamefont {E.}~\bibnamefont {Ferrara}}, \bibinfo
  {author} {\bibfnamefont {K.}~\bibnamefont {McKelvey}}, \bibinfo {author}
  {\bibfnamefont {F.}~\bibnamefont {Menczer}}, \ and\ \bibinfo {author}
  {\bibfnamefont {A.}~\bibnamefont {Flammini}},\ }\href {\doibase
  10.1371/journal.pone.0055957} {\bibfield  {journal} {\bibinfo  {journal}
  {PLoS ONE}\ }\textbf {\bibinfo {volume} {8}},\ \bibinfo {pages} {e55957}
  (\bibinfo {year} {2013})}\BibitemShut {NoStop}%
\bibitem [{\citenamefont {Fortunato}\ and\ \citenamefont
  {Barth\'{e}lemy}(2007)}]{Fortunato2007}%
  \BibitemOpen
  \bibfield  {author} {\bibinfo {author} {\bibfnamefont {S.}~\bibnamefont
  {Fortunato}}\ and\ \bibinfo {author} {\bibfnamefont {M.}~\bibnamefont
  {Barth\'{e}lemy}},\ }\href {\doibase 10.1073/pnas.0605965104} {\bibfield
  {journal} {\bibinfo  {journal} {Proceedings of the National Academy of
  Sciences of the United States of America}\ }\textbf {\bibinfo {volume}
  {104}},\ \bibinfo {pages} {36} (\bibinfo {year} {2007})}\BibitemShut
  {NoStop}%
\bibitem [{\citenamefont {Schaub}\ \emph {et~al.}(2012)\citenamefont {Schaub},
  \citenamefont {Delvenne}, \citenamefont {Yaliraki},\ and\ \citenamefont
  {Barahona}}]{Schaub2012}%
  \BibitemOpen
  \bibfield  {author} {\bibinfo {author} {\bibfnamefont {M.~T.}\ \bibnamefont
  {Schaub}}, \bibinfo {author} {\bibfnamefont {J.-C.}\ \bibnamefont
  {Delvenne}}, \bibinfo {author} {\bibfnamefont {S.~N.}\ \bibnamefont
  {Yaliraki}}, \ and\ \bibinfo {author} {\bibfnamefont {M.}~\bibnamefont
  {Barahona}},\ }\href {\doibase 10.1371} {\bibfield  {journal} {\bibinfo
  {journal} {PLoS ONE}\ ,\ \bibinfo {pages} {1}} (\bibinfo {year}
  {2012})}\BibitemShut {NoStop}%
\bibitem [{\citenamefont {Traag}\ \emph {et~al.}(2011)\citenamefont {Traag},
  \citenamefont {{Van Dooren}},\ and\ \citenamefont {Nesterov}}]{Traag2011}%
  \BibitemOpen
  \bibfield  {author} {\bibinfo {author} {\bibfnamefont {V.~A.}\ \bibnamefont
  {Traag}}, \bibinfo {author} {\bibfnamefont {P.}~\bibnamefont {{Van Dooren}}},
  \ and\ \bibinfo {author} {\bibfnamefont {Y.}~\bibnamefont {Nesterov}},\
  }\href {http://arxiv.org/abs/1104.3083} {\bibfield  {journal} {\bibinfo
  {journal} {Physical Review}\ }\textbf {\bibinfo {volume} {84}},\ \bibinfo
  {pages} {016114} (\bibinfo {year} {2011})}\BibitemShut {NoStop}%
\bibitem [{\citenamefont {Delvenne}\ \emph {et~al.}(2010)\citenamefont
  {Delvenne}, \citenamefont {Yaliraki},\ and\ \citenamefont
  {Barahona}}]{Delvenne2010}%
  \BibitemOpen
  \bibfield  {author} {\bibinfo {author} {\bibfnamefont {J.-C.}\ \bibnamefont
  {Delvenne}}, \bibinfo {author} {\bibfnamefont {S.~N.}\ \bibnamefont
  {Yaliraki}}, \ and\ \bibinfo {author} {\bibfnamefont {M.}~\bibnamefont
  {Barahona}},\ }\href
  {http://www.pubmedcentral.nih.gov/articlerender.fcgi?artid=2919907\&tool=pmcentrez\&rendertype=abstract}
  {\bibfield  {journal} {\bibinfo  {journal} {Proceedings of the National
  Academy of Sciences of the United States of America}\ }\textbf {\bibinfo
  {volume} {107}},\ \bibinfo {pages} {12755} (\bibinfo {year}
  {2010})}\BibitemShut {NoStop}%
\bibitem [{\citenamefont {Lambiotte}\ \emph {et~al.}(2008)\citenamefont
  {Lambiotte}, \citenamefont {Delvenne},\ and\ \citenamefont
  {Barahona}}]{Lambiotte2008}%
  \BibitemOpen
  \bibfield  {author} {\bibinfo {author} {\bibfnamefont {R.}~\bibnamefont
  {Lambiotte}}, \bibinfo {author} {\bibfnamefont {J.~C.}\ \bibnamefont
  {Delvenne}}, \ and\ \bibinfo {author} {\bibfnamefont {M.}~\bibnamefont
  {Barahona}},\ }\href {http://arxiv.org/abs/0812.1770} {\bibfield  {journal}
  {\bibinfo  {journal} {Arxiv preprint arXiv:0812.1770}\ }\textbf {\bibinfo
  {volume} {812}},\ \bibinfo {pages} {1} (\bibinfo {year} {2008})}\BibitemShut
  {NoStop}%
\bibitem [{\citenamefont {Traag}\ \emph {et~al.}(2013)\citenamefont {Traag},
  \citenamefont {Krings},\ and\ \citenamefont {Van~Dooren}}]{Traag2013}%
  \BibitemOpen
  \bibfield  {author} {\bibinfo {author} {\bibfnamefont {V.}~\bibnamefont
  {Traag}}, \bibinfo {author} {\bibfnamefont {G.}~\bibnamefont {Krings}}, \
  and\ \bibinfo {author} {\bibfnamefont {P.}~\bibnamefont {Van~Dooren}},\
  }\href@noop {} {\  (\bibinfo {year} {2013})},\ \Eprint
  {http://arxiv.org/abs/1306.3398} {arXiv:1306.3398} \BibitemShut {NoStop}%
\bibitem [{\citenamefont {Rosvall}\ and\ \citenamefont
  {Bergstrom}(2008)}]{Rosvall2008}%
  \BibitemOpen
  \bibfield  {author} {\bibinfo {author} {\bibfnamefont {M.}~\bibnamefont
  {Rosvall}}\ and\ \bibinfo {author} {\bibfnamefont {C.~T.}\ \bibnamefont
  {Bergstrom}},\ }\href {http://arxiv.org/abs/0707.0609} {\bibfield  {journal}
  {\bibinfo  {journal} {Proceedings of the National Academy of Sciences of the
  United States of America}\ }\textbf {\bibinfo {volume} {105}},\ \bibinfo
  {pages} {1118} (\bibinfo {year} {2008})}\BibitemShut {NoStop}%
\bibitem [{\citenamefont {Fortunato}(2010)}]{Fortunato2010}%
  \BibitemOpen
  \bibfield  {author} {\bibinfo {author} {\bibfnamefont {S.}~\bibnamefont
  {Fortunato}},\ }\href {\doibase 10.1016/j.physrep.2009.11.002} {\bibfield
  {journal} {\bibinfo  {journal} {Physics Reports}\ }\textbf {\bibinfo {volume}
  {486}},\ \bibinfo {pages} {75} (\bibinfo {year} {2010})}\BibitemShut
  {NoStop}%
\bibitem [{\citenamefont {Porter}\ \emph {et~al.}(2009)\citenamefont {Porter},
  \citenamefont {Onnela},\ and\ \citenamefont {Mucha}}]{Porter2009}%
  \BibitemOpen
  \bibfield  {author} {\bibinfo {author} {\bibfnamefont {M.}~\bibnamefont
  {Porter}}, \bibinfo {author} {\bibfnamefont {J.}~\bibnamefont {Onnela}}, \
  and\ \bibinfo {author} {\bibfnamefont {P.}~\bibnamefont {Mucha}},\ }\href
  {http://www.ams.org/notices/200909/rtx090901082p.pdf} {\bibfield  {journal}
  {\bibinfo  {journal} {Notices of the AMS}\ }\textbf {\bibinfo {volume} {56}}
  (\bibinfo {year} {2009})}\BibitemShut {NoStop}%
\bibitem [{\citenamefont {Brandes}\ \emph {et~al.}(2008)\citenamefont
  {Brandes}, \citenamefont {Delling}, \citenamefont {Gaertler}, \citenamefont
  {Gorke}, \citenamefont {Hoefer}, \citenamefont {Nikoloski},\ and\
  \citenamefont {Wagner}}]{Brandes2008}%
  \BibitemOpen
  \bibfield  {author} {\bibinfo {author} {\bibfnamefont {U.}~\bibnamefont
  {Brandes}}, \bibinfo {author} {\bibfnamefont {D.}~\bibnamefont {Delling}},
  \bibinfo {author} {\bibfnamefont {M.}~\bibnamefont {Gaertler}}, \bibinfo
  {author} {\bibfnamefont {R.}~\bibnamefont {Gorke}}, \bibinfo {author}
  {\bibfnamefont {M.}~\bibnamefont {Hoefer}}, \bibinfo {author} {\bibfnamefont
  {Z.}~\bibnamefont {Nikoloski}}, \ and\ \bibinfo {author} {\bibfnamefont
  {D.}~\bibnamefont {Wagner}},\ }\href {\doibase 10.1109/TKDE.2007.190689}
  {\bibfield  {journal} {\bibinfo  {journal} {IEEE Transactions on Knowledge
  and Data Engineering}\ }\textbf {\bibinfo {volume} {20}},\ \bibinfo {pages}
  {172} (\bibinfo {year} {2008})}\BibitemShut {NoStop}%
\bibitem [{\citenamefont {Blondel}\ \emph {et~al.}(2008)\citenamefont
  {Blondel}, \citenamefont {Guillaume}, \citenamefont {Lambiotte},\ and\
  \citenamefont {Lefebvre}}]{Blondel2008}%
  \BibitemOpen
  \bibfield  {author} {\bibinfo {author} {\bibfnamefont {V.~D.}\ \bibnamefont
  {Blondel}}, \bibinfo {author} {\bibfnamefont {J.-L.}\ \bibnamefont
  {Guillaume}}, \bibinfo {author} {\bibfnamefont {R.}~\bibnamefont
  {Lambiotte}}, \ and\ \bibinfo {author} {\bibfnamefont {E.}~\bibnamefont
  {Lefebvre}},\ }\href {\doibase 10.1088/1742-5468/2008/10/P10008} {\bibfield
  {journal} {\bibinfo  {journal} {Journal of Statistical Mechanics: Theory and
  Experiment}\ }\textbf {\bibinfo {volume} {2008}},\ \bibinfo {pages} {P10008}
  (\bibinfo {year} {2008})}\BibitemShut {NoStop}%
\bibitem [{Note1()}]{Note1}%
  \BibitemOpen
  \bibinfo {note} {Code available at
  http://sites.uclouvain.be/absil/browet}\BibitemShut {NoStop}%
\bibitem [{\citenamefont {West}(2001)}]{West2001}%
  \BibitemOpen
  \bibfield  {author} {\bibinfo {author} {\bibfnamefont {D.~B.}\ \bibnamefont
  {West}},\ }\href
  {http://imamat.oxfordjournals.org/content/24/3/local/back-matter.pdf} {\emph
  {\bibinfo {title} {{Introduction to Graph Theory}}}},\ Dover books on
  advanced mathematics\ (\bibinfo  {publisher} {Prentice Hall},\ \bibinfo
  {year} {2001})\BibitemShut {NoStop}%
\bibitem [{\citenamefont {Sarlette}\ \emph {et~al.}(2008)\citenamefont
  {Sarlette}, \citenamefont {Tuna}, \citenamefont {Blondel},\ and\
  \citenamefont {Sepulchre}}]{sarlette2008global}%
  \BibitemOpen
  \bibfield  {author} {\bibinfo {author} {\bibfnamefont {A.}~\bibnamefont
  {Sarlette}}, \bibinfo {author} {\bibfnamefont {S.~E.}\ \bibnamefont {Tuna}},
  \bibinfo {author} {\bibfnamefont {V.}~\bibnamefont {Blondel}}, \ and\
  \bibinfo {author} {\bibfnamefont {R.}~\bibnamefont {Sepulchre}},\ }in\ \href
  {http://orbi.ulg.ac.be/jspui/bitstream/2268/9541/2/ifac2.pdf} {\emph
  {\bibinfo {booktitle} {Proceedings of the 17th IFAC World Congress}}}\
  (\bibinfo {year} {2008})\BibitemShut {NoStop}%
\bibitem [{\citenamefont {Carli}\ \emph {et~al.}(2010)\citenamefont {Carli},
  \citenamefont {Fagnani}, \citenamefont {Frasca},\ and\ \citenamefont
  {Zampieri}}]{carli2010gossip}%
  \BibitemOpen
  \bibfield  {author} {\bibinfo {author} {\bibfnamefont {R.}~\bibnamefont
  {Carli}}, \bibinfo {author} {\bibfnamefont {F.}~\bibnamefont {Fagnani}},
  \bibinfo {author} {\bibfnamefont {P.}~\bibnamefont {Frasca}}, \ and\ \bibinfo
  {author} {\bibfnamefont {S.}~\bibnamefont {Zampieri}},\ }\href
  {http://arxiv.org/pdf/0907.0748} {\bibfield  {journal} {\bibinfo  {journal}
  {Automatica}\ }\textbf {\bibinfo {volume} {46}},\ \bibinfo {pages} {70}
  (\bibinfo {year} {2010})}\BibitemShut {NoStop}%
\bibitem [{\citenamefont {Clauset}\ \emph {et~al.}(2004)\citenamefont
  {Clauset}, \citenamefont {Newman},\ and\ \citenamefont
  {Moore}}]{Clauset2004}%
  \BibitemOpen
  \bibfield  {author} {\bibinfo {author} {\bibfnamefont {A.}~\bibnamefont
  {Clauset}}, \bibinfo {author} {\bibfnamefont {M.~E.~J.}\ \bibnamefont
  {Newman}}, \ and\ \bibinfo {author} {\bibfnamefont {C.}~\bibnamefont
  {Moore}},\ }\href {http://arxiv.org/abs/cond-mat/0408187} {\bibfield
  {journal} {\bibinfo  {journal} {Physical Review E}\ }\textbf {\bibinfo
  {volume} {70}},\ \bibinfo {pages} {1} (\bibinfo {year} {2004})}\BibitemShut
  {NoStop}%
\bibitem [{\citenamefont {Rosvall}\ and\ \citenamefont
  {Bergstrom}(2011)}]{Rosvall2011}%
  \BibitemOpen
  \bibfield  {author} {\bibinfo {author} {\bibfnamefont {M.}~\bibnamefont
  {Rosvall}}\ and\ \bibinfo {author} {\bibfnamefont {C.~T.}\ \bibnamefont
  {Bergstrom}},\ }\href {http://arxiv.org/abs/1010.0431} {\bibfield  {journal}
  {\bibinfo  {journal} {PLoS ONE}\ }\textbf {\bibinfo {volume} {6}},\ \bibinfo
  {pages} {10} (\bibinfo {year} {2011})}\BibitemShut {NoStop}%
\bibitem [{\citenamefont {Schuetz}\ and\ \citenamefont
  {Caflisch}(2008)}]{Schuetz2008}%
  \BibitemOpen
  \bibfield  {author} {\bibinfo {author} {\bibfnamefont {P.}~\bibnamefont
  {Schuetz}}\ and\ \bibinfo {author} {\bibfnamefont {A.}~\bibnamefont
  {Caflisch}},\ }\href {http://arxiv.org/abs/0712.1163} {\bibfield  {journal}
  {\bibinfo  {journal} {Physical Review E - Statistical, Nonlinear and Soft
  Matter Physics}\ }\textbf {\bibinfo {volume} {77}},\ \bibinfo {pages}
  {046112} (\bibinfo {year} {2008})}\BibitemShut {NoStop}%
\bibitem [{\citenamefont {Raghavan}\ \emph {et~al.}(2007)\citenamefont
  {Raghavan}, \citenamefont {Albert},\ and\ \citenamefont
  {Kumara}}]{Raghavan2007}%
  \BibitemOpen
  \bibfield  {author} {\bibinfo {author} {\bibfnamefont {U.~N.}\ \bibnamefont
  {Raghavan}}, \bibinfo {author} {\bibfnamefont {R.}~\bibnamefont {Albert}}, \
  and\ \bibinfo {author} {\bibfnamefont {S.}~\bibnamefont {Kumara}},\ }\href
  {http://arxiv.org/abs/0709.2938} {\bibfield  {journal} {\bibinfo  {journal}
  {Physical Review E - Statistical, Nonlinear and Soft Matter Physics}\
  }\textbf {\bibinfo {volume} {76}},\ \bibinfo {pages} {036106} (\bibinfo
  {year} {2007})}\BibitemShut {NoStop}%
\bibitem [{\citenamefont {Csardi}\ and\ \citenamefont
  {Nepusz}(2006)}]{Csardi2006}%
  \BibitemOpen
  \bibfield  {author} {\bibinfo {author} {\bibfnamefont {G.}~\bibnamefont
  {Csardi}}\ and\ \bibinfo {author} {\bibfnamefont {T.}~\bibnamefont
  {Nepusz}},\ }\href {http://igraph.sf.net} {\bibfield  {journal} {\bibinfo
  {journal} {InterJournal}\ }\textbf {\bibinfo {volume} {Complex Systems}},\
  \bibinfo {pages} {1695} (\bibinfo {year} {2006})}\BibitemShut {NoStop}%
\bibitem [{\citenamefont {Lancichinetti}\ and\ \citenamefont
  {Fortunato}(2009)}]{Lancichinetti2009}%
  \BibitemOpen
  \bibfield  {author} {\bibinfo {author} {\bibfnamefont {A.}~\bibnamefont
  {Lancichinetti}}\ and\ \bibinfo {author} {\bibfnamefont {S.}~\bibnamefont
  {Fortunato}},\ }\href {http://arxiv.org/abs/0904.3940} {\bibfield  {journal}
  {\bibinfo  {journal} {Physical Review E}\ }\textbf {\bibinfo {volume} {80}},\
  \bibinfo {pages} {9} (\bibinfo {year} {2009})}\BibitemShut {NoStop}%
\bibitem [{\citenamefont {Danon}\ \emph {et~al.}(2005)\citenamefont {Danon},
  \citenamefont {Duch}, \citenamefont {Diaz-Guilera},\ and\ \citenamefont
  {Arenas}}]{Danon2005}%
  \BibitemOpen
  \bibfield  {author} {\bibinfo {author} {\bibfnamefont {L.}~\bibnamefont
  {Danon}}, \bibinfo {author} {\bibfnamefont {J.}~\bibnamefont {Duch}},
  \bibinfo {author} {\bibfnamefont {A.}~\bibnamefont {Diaz-Guilera}}, \ and\
  \bibinfo {author} {\bibfnamefont {A.}~\bibnamefont {Arenas}},\ }\href
  {http://arxiv.org/abs/cond-mat/0505245} {\bibfield  {journal} {\bibinfo
  {journal} {Journal of Statistical Mechanics: Theory and Experiment}\ }\textbf
  {\bibinfo {volume} {2005}},\ \bibinfo {pages} {10} (\bibinfo {year}
  {2005})}\BibitemShut {NoStop}%
\end{thebibliography}%

\end{document}